\documentclass[journal=jctcce,manuscript=article,layout=twocolumn]{achemso}
\usepackage{graphicx}
\usepackage{amsmath}
\usepackage{wrapfig}
\usepackage{physics}
\usepackage{caption}
\usepackage{subfiles}

\usepackage{array}
\usepackage{ragged2e}
\newcolumntype{P}[1]{>{\RaggedRight\hspace{0pt}}p{#1}}

\graphicspath{ {Figures/} }

\title{A Thermal Gradient Approach for the Quasi-Harmonic Approximation and its Application to Improved Treatment of
Anisotropic Expansion}

\author{Nathan S. Abraham}
\affiliation{Department of Chemical and Biological Engineering, University of Colorado Boulder, Boulder, CO 80309, USA}
\author{Michael R. Shirts}
\affiliation{Department of Chemical and Biological Engineering, University of Colorado Boulder, Boulder, CO 80309, USA}
\email{michael.shirts@colorado.edu}

\begin{document}
\maketitle
\bibliographystyle{plain}
\begin{abstract}
    We present a novel approach to efficiently implement thermal expansion in the quasi-harmonic approximation (QHA) 
for both isotropic and more importantly, anisotropic expansion. In this approach, we rapidly determine a crystal's 
equilibrium volume and shape at a given temperature by integrating along the gradient of expansion from zero Kelvin up 
to the desired temperature. We compare our approach to previous isotropic methods that rely on a brute-force grid search 
to determine the free energy minimum, which is infeasible to carry out for anisotropic expansion, as well as 
quasi-anisotropic approaches that take into account the contributions to anisotropic expansion from the lattice energy.
We compare these methods for experimentally known polymorphs of piracetam and resorcinol and show that both isotropic methods agree 
to within error up to 300 K. Using the Gr\"{u}neisen parameter causes up to 0.04 kcal/mol deviation in the Gibbs free
energy, but for polymorph free energy differences there is a cancellation in error with all isotropic methods 
within 0.025 kcal/mol at 300 K.

    Anisotropic expansion allows the crystals to relax into lattice geometries 0.01--0.23 kcal/mol lower in energy at
300 K relative to isotropic expansion. For polymorph free energy differences all QHA methods produced results 
within 0.02 kcal/mol of each other for resorcinol and 0.12 kcal/mol for piracetam, the two molecules tested here, 
demonstrating a cancellation of error for isotropic methods. 

    We also find that when expanding in more than a single volume 
variable, there is a non-negligible rate of failure of the basic approximations of QHA. Specifically, 
while expanding into new harmonic modes as the box vectors are increased, the system often falls into alternate, 
structurally distinct harmonic modes unrelated by continuous deformation from the original harmonic mode.
\end{abstract}

\section{Introduction}
    Computational screening of organic solid materials has garnered significant additional interest within the past 
decade. The most notable contributions to this field have emerged through crystal structure prediction (CSP) algorithms. 
CSPs provide a way to find potential polymorphs that can vary significantly in properties from previously known 
structures.~\cite{chemburkar_dealing_2000,bauer_ritonavir:_nodate,singhal_drug_2004,haas_high_2007,giri_tuning_2011}. 
CSPs use a combination of exhaustive searches of crystal lattice packings and carefully tuned force fields adapted to 
solid state organic molecules to accurately predict stable crystal conformations. They have shown significant progress 
over the last decade in identifying experimentally observed polymorphs, or at least ranking experimental structures 
relatively highly among many alternate structures, in a series of blind prediction 
challenges. Over the years, the predictions have been moving towards crystals composed of larger and more flexible 
molecules~\cite{reilly_report_2016,motherwell_crystal_2002,lommerse_test_2000,day_third_2005,day_significant_2009,bardwell_towards_2011}.

    A current limitation in CSP approaches is the static treatment of the 
crystal.~\cite{price_computed_2009} Typically, CSPs rely on the lattice energy to rank predicted structures, which 
excludes any dynamic behavior in the crystal. One can improve the CSP estimate by adding the entropy due to lattice 
vibrations. In the simplest lattice dynamic model, the harmonic approximation (HA), we assume that the crystal vibrates 
as a set of harmonic oscillators, providing a harmonic approximation to the Gibbs free energy:
  \begin{eqnarray}
    G(T) = \min_{\boldsymbol{x}} \left(U(V_0,\boldsymbol{x})\right) + A_{v}(V_0,T) + PV
    \label{equation:HA_G}
  \end{eqnarray}
In the HA, we assume the Gibbs free energy ($G$) is a function of temperature and is a simple sum of the lattice energy ($U$) and the 
temperature dependent harmonic vibrational free energy ($A_{v}$) of the given structure. In eq.~\ref{equation:HA_G}, $T$ is the temperature
of interest, $\min_{\boldsymbol{x}} \left(U(V_0,\boldsymbol{x})\right)$ is the lattice energy ($U$) at the lattice minimum volume 
($V_{0}$) and geometry minimized coordinates ($\boldsymbol{x}$), and $P$ is the pressure. Nyman and Day showed that 9\% 
of 500 pairs of small molecule polymorphs re-ranked between 0 and 300 K.~\cite{nyman_static_2015} This percentage is 
still significant, considering that thermal expansion of the crystal is neglected in this estimate.

    To most accurately model crystals, one should include the thermal expansion of a crystal. 
Crystals expand with temperature, increasing the total potential energy of the crystal, but also increasing the 
entropy. To improve upon the lattice dynamic model, we can use the quasi-harmonic approximation (QHA), which corrects 
eq.~\ref{equation:HA_G} by treating it as an argument-minimization of the volume:
  \begin{eqnarray}
    G(T) = \min_{V} f(V,T) \\
    f(V,T) = \min_{\boldsymbol{x}} \left(U(V,\boldsymbol{x})\right) + A_{v}(V,T) + PV
    \label{equation:QHA_G_iso}
  \end{eqnarray}
While being more computationally demanding, this model provides a more realistic depiction of the way solid materials 
behave, especially at low temperatures where anharmonic vibrations are negligible. As a follow-up study to their 
previous work on harmonic analysis of a large set of polymorphs, Nyman and Day re-evaluated the same structures using 
QHA, and found that 21\% of the crystal structures re-ranked from 0 K to their melting 
temperature,~\cite{nyman_modelling_2016} further demonstrating that more accurate representation of finite temperature 
crystals can be important.

    In eq.~\ref{equation:QHA_G_iso} it is most straightforward and most common to perform the argument minimization 
with an isotropic constraint on the lattice volume, holding lattice vectors proportional and angles 
fixed. While isotropic QHA methods have worked for small organic crystals, Heit et al. discuss how  
these predictions could be improved by considering anisotropic expansion.~\cite{heit_predicting_2015} 
In particular, anisotropic treatments are expected to become more prevalent for larger molecules or crystal structures that 
are inherently layered,~\cite{boldyreva_anisotropic_2000,varughese_interaction_2011,dolan_structures_2008,wada_lateral_2002,bolotina_energetic_2004} 
which is characteristic of many potential drug candidates or for use in organic semiconductors.

    In a recent study, we examined the Gibbs free energy differences as a function of temperature between polymorphs of 
12 small organic molecules with both molecular dynamics (MD) and QHA using isotropic expansion. We found that for
the small, rigid, ordered crystals QHA and MD agreed within $<$ 0.07 kcal/mol, whereas for the larger, flexible 
molecules or disordered crystals QHA deviated up to 0.37 kcal/mol from MD at 300 K. Because we were using an 
isotropic thermal expansion model for QHA we could not determine if the deviations between the QHA and MD were due to 
anharmonic motions, neglected in the QHA, or due to anisotropic thermal expansion. While MD showed that all 24 crystals 
exhibited some level of anisotropy during thermal expansion, the degree of anisotropy was greater for those systems in
which QHA and MD deviated $>0.1$ kcal/mol for polymorph free energy differences,~\cite{dybeck_comparison_2016} motivating 
a closer look at the validity of the isotropic expansion approximation.

    The standard approach for performing isotropic QHA is not directly amenable to a fully anisotropic free energy minimization.
This standard approach requires a stepwise expansion of the lattice minimum structure to a set of new volumes, $V_{i}$, and 
re-minimizing the coordinates for each new volume. For each of these volumes, $V_{i}$, the harmonic approximation is 
calculated at each volume. At each temperature of interest, we then determine which of these HA calculations 
corresponds to the volume that minimizes $G$.~\cite{ramirez_quasi-harmonic_2012-1} We describe the procedure in more 
detail below. In theory, this approach could be applied to anisotropic expansion by varying all six box vector 
parameters, forming a large grid of lattice structures of different shapes and sizes.~\cite{filippini_lattice-dynamical_1975,filippini_deriving_1981} 
However, if we consider all six dimensions of anisotropic expansion, this approach scales with $N^{6}$ where N is the 
number of grid steps in each dimension. To converge on the true thermodynamic minimum the steps between expansions must 
be small, which is particularly problematic with scaling to the 6th power in the number of steps. One could fit some 
multidimensional function to a sparser grid, but there are no existing explicitly anisotropic equations of state, 
and other multidimensional splines are complex and may not converge similarly in differently behaving systems.

    There have been important recent developments towards the inclusion of anisotropy in QHA calculations that attempt 
to get around these scaling problems. Erba et al. developed an approach that performs a stepwise expansion of the 
crystal lattice,~\cite{erba_combining_2014,dovesi_quantum-mechanical_2018} but also includes an additional step where 
the compressed and expanded structures are energy minimized at the fixed expanded volume. To avoid too fine a grid of 
expansion and compression, the lattice energy is fit to an equation of state (in volume alone) and the vibrational 
frequencies, determined at these optimized structures, are fit with a polynomial (again in volume).  With these two 
equations, the properties using eq.~\ref{equation:QHA_G_iso} can quickly be fit at any temperature or pressure, 
assuming the EoS is sufficiently accurate. This approach has been used in conjunction with \textit{ab initio} 
techniques for a number of materials and organic crystals, which has shown notable success matching experimental 
results up to high temperatures and 
pressures.~\cite{erba_structural_2015,erba_thermal_2016,erba_how_2015,erba_assessing_2015,maul_thermal_2016}
However, this approach performs the cell optimization based on the lattice energy alone, meaning the lattice
geometry will not be at free energy minimum for that volume. 

   A nearly equivalent variation of this approach performs lattice minimization at a number of different increased and 
decreased pressures.~\cite{cervinka_ab_2017} Similar to Erba's method, these $(U,V)$ pairs, which implicitly include 
anisotropic expansion due to the lattice energy, are fit to an equation of state. These structures are minimum lattice energy 
structures at their volume; the only difference to the direct volume-constrained minimization is that these structures
are likely not at uniform volume intervals, as the pressure is specified, not the volume.

    Other options to employ anisotropic expansion for QHA, but most have drawbacks due to expense or are not at the 
true quasi-harmonic minimum. Some approaches use prior experimental knowledge,~\cite{day_nonempirical_2003} which are 
ineffective for crystal structure predictions where high temperature lattice geometries are unknown. 
Alternatively, one could directly optimize the free energy with respect to all lattice parameters at each 
temperature~\cite{taylor_free-energy_1997,taylor_quasiharmonic_1999}; however, this requires a full nonlinear 
optimization of the sum of the lattice energy and the vibrational energy at each state point of interest, which 
typically involves a number of approximation for the vibrational energy~\cite{taylor_free-energy_1997}. This approach 
results in significant additional computational work, often requiring new derivations depending on the energy model used, 
which cannot be reused between state points without further approximation. 

    In this paper, we present a new way to efficiently compute the free energies of crystalline solids in the QHA for 
both isotropic and anisotropic thermal expansion based on estimating the gradient of thermal expansion. By computing
the gradient of thermal expansion we are only limited by the numerical integration used and inherent failures in the 
QHA, which we will discuss in this paper. We compare our method to the stepwise approach for isotropic QHA 
and what has been referred to~\cite{cervinka_ab_2017} as the ``quasi-anisotropic'' method of Erba and 
others.~\cite{crystal_manual,erba_combining_2014,cervinka_ab_2017} We examine how the free energies, the free energy 
differences between polymorphs, and the crystals geometries vary across the different methods. We also show how 
those thermodynamic properties vary when using the Gr\"{u}neisen parameter to account for changes in the crystals 
vibrational spectra.

\section{Theory}
\subsection{Harmonic Approximation}
    In the simplest lattice dynamic model, the harmonic approximation (HA), we assume that thermal expansion is 
negligible. The Gibbs free energy for HA is shown in eq.~\ref{equation:HA_G}, where $\min_{x} \left(U(V_0,\boldsymbol{x})\right)$ 
is the minimum of the internal energy at the crystal lattice minimum energy volume ($V_0$). The vibrational free energy,
$A_{v}(V,T)$, is calculated as:
  \begin{eqnarray}
    A_{v}(V,T) = \sum_{k} \frac{1}{\beta} \ln{\left(\beta \hbar \omega_{k}(V)\right)}
    \label{equation:QHA_Vib}
  \end{eqnarray}
Where $\beta$ is $\left(k_{B}T\right)^{-1}$ and $\omega_{k}(V)$ is the vibrational frequency of the $k^{th}$ vibrational mode. Since 
the kinetic energy will be the same between polymorphs, we can simply use the potential energy instead of the total 
energy when examining the free energy difference between polymorphs. Note that the above formula is the classical 
approximation, though the approach presented here will be equally valid for the quantum harmonic oscillator formula.

\subsection{Quasi-Harmonic Approximation} \label{section:Quasi-Harmonic_Approximation}
    The quasi-harmonic approximation (QHA) more accurately captures entropic effects compared to HA by accounting for 
the entropic effects caused by thermal expansion. There are two ways to account for thermal expansion: isotropic and 
anisotropic. Isotropic thermal expansion assumes that 1) the lattice vectors remain proportional for all volumes and 
2) that the lattice angles remain fixed. Isotropic expansion is a computationally fast approximation to the true 
thermal expansion, and has been shown to work well for number of 
materials,~\cite{heit_predicting_2015,ramirez_quasi-harmonic_2012-1,winkler_thermodynamic_1992} but can constrain the 
crystal to the wrong lattice geometry at high temperatures for materials that expand
anisotropically.~\cite{iwanaga_anisotropic_2000,das_exceptionally_2010} In isotropic QHA, we calculate the minimum of 
the harmonic approximation over all volumes at a given temperature $T$. At $T=0$, this will coincide with the harmonic 
approximation, but at $T>0$ QHA will minimize $G$ to a lower free energy minimum and 
larger volume. 

    In contrast, anisotropic expansion allows the 6 lattice parameters to change with no constraints. This is a more
accurate model to describe thermal expansion, since most crystals exhibit some level of anisotropy. For this more accurate
expansion model, we must minimize the argument over the crystal lattice tensor $\textbf{C}$, which defines the box shape of
a single unit cell:
  \begin{eqnarray}
    \boldsymbol{C} = 
     \begin{bmatrix}
     C_{1} & C_{4} & C_{5} \\
     0     & C_{2} & C_{6} \\
     0     & 0     & C_{3} 
     \end{bmatrix}
  \end{eqnarray}
As the energy of the crystal is a function of six lattice parameters, $C_{i}$ for $i = 1, 2, ... 6$, not simply the 
volume alone. 
  \begin{eqnarray}
    G(T) = \min_{\textbf{C}} f(\textbf{C},T) \\
    f(\textbf{C},T) = \min_{x} \left(U(\textbf{C},x)\right) + A_{v}(\textbf{C},T) + PV(\textbf{C})
    \label{equation:QHA_G_aniso}
  \end{eqnarray}
Full anisotropic expansion is difficult to carry out, because it requires optimization of a complicated argument in a 
substantially larger space.

\subsection{Gr\"{u}neisen Parameter} \label{section:Gruneisen_Parameter}
    The vibrational spectra of the crystal changes as the strains are applied to the system, which must be accounted 
for in QHA. There are two options to compute ${\omega_{k}}$ for all values of $\textbf{C}$. The first is to compute the 
mass-weighted Hessian for every structure with a unique lattice tensor $\textbf{C}$, which is the most expensive step in the approximation. 
The second option is to assume that the changes in $\omega_{k}$ are proportional to the strains applied to the crystal, 
which is described by the Gr\"{u}neisen parameter. This approach is a standard technique used to speed up lattice 
dynamics calculations~\cite{heit_predicting_2015,ramirez_quasi-harmonic_2012-1,nath_high-throughput_2016,huang_efficient_2016}.

\subsubsection{Isotropic Gr\"{u}neisen Parameter}
    For isotropic thermal expansion, the Gr\"{u}neisen parameter of a particular mode is the proportional change between 
that vibrational frequency and the lattice volume. To calculate this and minimize rounding error, we take the alternative 
formulation:
  \begin{eqnarray}
    \frac{\partial V}{V}  =  -\frac{1}{\gamma_{k}} \frac{\partial \omega_{k}(V)}{\omega_{k}(V)}
    \label{equation:Gru_Iso_proportional}
  \end{eqnarray}
Where again $V$ is the isotropic volume, $\omega_{k} (V)$ is the vibrational frequency of the $k^{th}$ mode at $V$, and 
$\gamma_{k}$ is the Gr\"{u}neisen parameter of the $k^{th}$ mode. 

    By computing the derivatives of the log quantities by forward finite difference and rearranging 
eq.~\ref{equation:Gru_Iso_proportional} we get eq.~\ref{equation:Gru_Iso_solution}. We use the forward difference approach rather 
than the central difference approach because we require the 0 K vibrational spectra in any case, so computing a central difference will be 
at least 50\% more expensive. Eq.~\ref{equation:Gru_Iso_proportional} can be integrated to get 
eq.~\ref{equation:Gru_Iso_new} and the solution from eq.~\ref{equation:Gru_Iso_solution} can be used to determine the 
$k^{th}$ vibrational frequency at any subsequent isotropically expanded structure. 
  \begin{eqnarray}
    \gamma_{k} = -\frac{\ln\left({\omega_{k}(V + \Delta V)}\right) - \ln\left({\omega_{k}(V)}\right)}{\ln\left({1 + \frac{\Delta V}{V}}\right)} \label{equation:Gru_Iso_solution} \\
    \omega_{k}(V + \Delta V) = \omega_{k}(V) \left(1 + \frac{\Delta V}{V}\right)^{-\gamma_{k}}
    \label{equation:Gru_Iso_new}
  \end{eqnarray}
Where $\omega_{k}(V)$ is the $k^{th}$ vibrational frequency for the lattice minimum structure, $V$ is the lattice minimum 
volume, $V + \Delta V$ is the volume of the isotropically expanded structure, and $\omega_{k}(V + \Delta V)$ is the 
$k^{th}$ vibrational frequency for the expanded structure.
  
\subsubsection{Anisotropic Gr\"{u}neisen Parameter} \label{section:Aniso_GRU}
    We can also assume that the change in relative vibrational frequencies is a linear combination of changes in each 
relative box vector, not just the volume itself, resulting in six anisotropic Gr\"{u}neisen parameters. To develop 
the formalism of inclusion of anisotropic expansion, we present the formalism from Choy et al.~\cite{choy_thermal_1984}.
Eq.~\ref{equation:Gru_Aniso_proportional} describes the proportionality of a strain ($\eta_{i}$) placed on the 
lattice on the $k^{th}$ vibrational frequency. Here $\eta_{i}$ is one of the six principle strains on the crystal lattice 
matrix, $i = 1, 2, ... 6$.
  \begin{eqnarray}
    \left(\partial \eta_{i}\right)_{\eta_{j} \ne \eta_{i}} = -\frac{\left(\partial \omega_{k}\right)_{\eta_{j} \ne \eta_{i}}}{\omega_{k} \gamma_{k,i}}
	\label{equation:Gru_Aniso_proportional}
  \end{eqnarray}
Similarly to the isotropic case, eq.~\ref{equation:Gru_Aniso_proportional} can be re-arranged and solved numerically for 
the six different strains shown in eq.~\ref{equation:Gru_Aniso_solution}.
  \begin{eqnarray}
    \gamma_{k,i} = - \frac{\ln\left({\omega_{k}(\eta_{i})}\right) - \ln\left({\omega_{k,0}}\right)}{\eta_{i}}
	\label{equation:Gru_Aniso_solution}
  \end{eqnarray}
    If we integrate eq.~\ref{equation:Gru_Aniso_proportional} and use the solution from 
eq.~\ref{equation:Gru_Aniso_solution}, than the change in the $k^{th}$ vibrational frequency due to strain $\eta_{i}$ can be 
described by eq.~\ref{equation:Gru_Aniso_new_strain}, where $\omega_{k}(\eta_{i})$ is the $k^{th}$ vibrational frequency for
the crystal under strain $\eta_{i}$, $\omega_{k,0}$ is the $k^{th}$ vibrational frequency of the lattice structure, and 
$\gamma_{k,i}$ is the Gr\"{u}neisen for the $k^{th}$ mode due to strain $\eta_{i}$.
  \begin{eqnarray}
    \omega_{k}(\eta_{i}) = \omega_{k,0} \exp\left({-\eta_{i} \gamma_{k,i}}\right)
	\label{equation:Gru_Aniso_new_strain}
  \end{eqnarray}  
Eq.~\ref{equation:Gru_Aniso_new_strain} can be used in a linear combination of all six strains on the system to predict 
the vibrational spectra due to any lattice strain. 
  \begin{eqnarray}
    \omega_{k}(\eta_{1,2, ... 6}) = \omega_{k,0} \exp\left({\sum_{i=1}^{6} -\eta_{i} \gamma_{k,i}}\right)
	\label{equation:Gru_Aniso_new_strain_full}
  \end{eqnarray}
    To compute the six principle strains placed on the crystal during thermal expansion requires us to relate the crystal
lattice tensor to strain. The six principle strains make up the strain tensor as:
  \begin{eqnarray}
    \boldsymbol{\eta} = 
     \begin{bmatrix}
     \eta_{1} & \eta_{4} & \eta_{5} \\
     \eta_{4} & \eta_{2} & \eta_{6} \\
     \eta_{5} & \eta_{6} & \eta_{3} 
     \end{bmatrix}
  \end{eqnarray}
For $\eta_{i}$, $i = 1,2$ and $3$ are normal strains and $i = 4,5$ and $6$ are shear strains, which are applied 
symmetrically. By assuming that the crystals are strained by small amounts and that angular momentum is conserved
we can relate the lattice minimum crystal tensor ($\boldsymbol{C_{0}}$) to a new expanded tensor ($\boldsymbol{C}$)
under strain $\boldsymbol{\eta}$, as:
  \begin{eqnarray}
    \boldsymbol{C} = \boldsymbol{C_{0}}\left(\boldsymbol{I} + \boldsymbol{\eta}\right)
  \end{eqnarray}
Re-arranging to determine the strain, we can compute the strain placed on the reference crystal, $\boldsymbol{C_{0}}$,
to produce some new crystal tensor, $\boldsymbol{C}$.
  \begin{eqnarray}
    \boldsymbol{\eta} = \boldsymbol{C} \boldsymbol{C_{0}}^{-1} - \boldsymbol{I}
	\label{equation:strain_from_CTensor}
  \end{eqnarray}
There is a artificial rotation caused in the computation of the strain in eq.~\ref{equation:strain_from_CTensor} due
to the representation of the crystal tensor. Since we are dealing with bulk crystals, any rotational strain computed
between two sets of lattice vectors will not correspond to a change in energy. Further discussion of the rotational
strain and how we remove it to compute $\boldsymbol{\eta}$ as a symmetric tensor can be found in the Supporting 
Information (Section~\ref{section:Computing_Strain}).

    Knowing the applied strain relative to the reference structure and the set of Gr\"{u}neisen parameters we can 
approximate the vibrational spectra of any crystal strained from the reference point by solving 
eq.~\ref{equation:Gru_Aniso_new_strain_full}.

\subsection{Stepwise Thermal Expansion}
    The standard approach for performing isotropic QHA works reasonably well for isotropic expansion, but is too 
inefficient for anisotropic expansion. Typically, the crystal lattice structure would be expanded and compressed to 
create a subset of structures that could be described by an array of isotropic volumes ${V_{i}}$. The vibrational 
spectra of each structure would be computed, allowing us to perform a harmonic approximation on each structure. At a 
given temperature, the isotropic volume, $V_{i}$, that minimizes the Gibbs free energy would be used to describe the 
system at $T$. This method allows us to perform QHA in a brute force manner for isotropic expansion, but becomes 
infeasible when considering anisotropic thermal expansion, which would require either a 6D grid search. A detailed 
description of stepwise isotropic QHA can be found in section~\ref{section:StepIso-QHA}.
 
\subsection{A New Approach: The Gradient Thermal Expansion}
    We wish to determine exactly how the structure expands with temperature to make isotropic expansion more efficient 
and provide a more suitable way to perform anisotropic expansion. We present an approach by Gould 
et al.~\cite{gould_differentiating_2016} to solve argument min- or maximization problems, and apply this method to 
determine the rate of thermal expansion of organic crystals. Eq.~\ref{equation:general_min} is a general minimization 
problem, which can be seen to be of the same class of problem as the equations for the QHA Gibbs free energy in 
eq.~\ref{equation:QHA_G_iso} and eq.~\ref{equation:QHA_G_aniso}. 
  \begin{eqnarray}
    g(t) = \min_{x} f(x,t)
	\label{equation:general_min}
  \end{eqnarray}
Where the solution to $g(t)$ is the equal to the minimum of $f(x,t)$ with respect to $x$. The gradient of $x$ as a 
function of $t$, in eq.~\ref{equation:general_gradient}, can be computed as (see Gould et al.)~\cite{gould_differentiating_2016}:
  \begin{eqnarray}
    \frac{\partial x}{\partial t} = - \frac{\frac{\partial^{2}g}{\partial x \partial t}}{\frac{\partial^{2} g}{\partial x^{2}}}
	\label{equation:general_gradient}
  \end{eqnarray}
If we can compute these partial derivatives, then we can use the gradient $\frac{dx}{dt}$ to solve the resulting 
ordinary differential equation for $x(t)$. Eq.~\ref{equation:general_gradient} can be applied to single or 
multidimensional problems, which we outline for both isotropic and anisotropic expansion. In these cases, the variable 
$t$ in eq.~\ref{equation:general_min} is the temperature $T$, and $x$ in eq.~\ref{equation:general_min} is either the 
volume $V$ or the components of the crystal tensor $\textbf{C}$.

\subsubsection{Gradient Isotropic Thermal Expansion}
    We can easily apply eq.~\ref{equation:general_min} to eq.~\ref{equation:QHA_G_iso} to obtain a thermal gradient 
approach to calculating the expansion needed for QHA expansions, obtaining for the volume:
  \begin{eqnarray}
    \frac{\partial V}{\partial T} = - \frac{\frac{\partial^{2}G}{\partial V \partial T}}{\frac{\partial^{2} G}{\partial V^{2}}} = \frac{\frac{\partial S}{\partial V}}{\frac{\partial^{2} G}{\partial V^{2}}}
	\label{equation:Gradient_Iso}
  \end{eqnarray}
We use the additional simplification of replacing the temperature derivative of $G$ with $-S$, as $S$ can be easily 
calculated analytically once $G$ is calculated. The analytical solution of $S$ is derived from eq.~\ref{equation:QHA_Vib}
and shown in eq.~\ref{equation:vib_S}.
  \begin{eqnarray}
    S(V,T) = \sum_{k} k_{B}\left(1 - \ln{\left(\beta \hbar \omega_{k}(V)\right)}\right)
	\label{equation:vib_S}
  \end{eqnarray}

    With the thermal gradient calculated in eq.~\ref{equation:Gradient_Iso}, we can determine how the minimum energy
volume changes with temperature by using any standard algorithm for ordinary differential equations; in this paper, 
we use the 4$^{th}$ order Runge-Kutta approach.

\subsubsection{Gradient Anisotropic Thermal Expansion} \label{section:Grad_Aniso}
    Eq.~\ref{equation:QHA_G_aniso} presents the thermal expansion gradient for anisotropic expansion; 
eq.~\ref{equation:Gradient_Aniso_orig} shows this solution using the thermal expansion approach.
  \begin{eqnarray}
    \frac{\partial \boldsymbol{C}}{\partial T} &=& - \left(\frac{\partial^{2} G}{\partial \boldsymbol{C}^{2}}\right)^{-1} \frac{\partial^{2}G}{\partial \boldsymbol{C} \partial T}  \\
&=& \left(\frac{\partial^{2} G}{\partial \boldsymbol{C}^{2}}\right)^{-1} \frac{\partial S}{\partial \boldsymbol{C}}
	\label{equation:Gradient_Aniso_orig}
  \end{eqnarray}
Here, $\boldsymbol{C}$ is the array of the three diagonal and three off diagonal elements, all of which describe the 
orientation of the lattice vectors:
  \begin{eqnarray}
    \boldsymbol{C} = 
     \begin{bmatrix}
     C_{1} & C_{2} & C_{3} & C_{4} & C_{5} & C_{6}
     \end{bmatrix}^{T}
  \end{eqnarray}
Given the dimensions of $\boldsymbol{C}$, the left hand side of eq.~\ref{equation:Gradient_Aniso_orig} and 
the second term on the right hand side are length-six vectors. The first term on the right hand side is a 6$\times$6 
matrix. Similar to the isotropic case, the anisotropic gradients are solved using a central finite difference approach.

    In some cases, parts of $\frac{\partial^{2} G}{\partial \boldsymbol{C}^{2}}$ due to a given lattice vector may be inherently zero, which will 
make the matrix singular. We therefore solve the thermal gradient using the implicit formulation in 
eq.~\ref{equation:Gradient_Aniso}.
  \begin{eqnarray}	
    \left(\frac{\partial^{2} G}{\partial \boldsymbol{C}^{2}}\right) \frac{\partial \boldsymbol{C}}{\partial T} &=& \frac{\partial S}{\partial \boldsymbol{C}}
   \label{equation:Gradient_Aniso}
  \end{eqnarray}

    Using the thermal gradient calculated in eq.~\ref{equation:Gradient_Aniso}, we can determine how the minimum energy 
volume changes with temperature by using any standard algorithm for, now, coupled first order ordinary differential 
equations; in this paper, we again use the Runge-Kutta approach.

\subsection{Constrained Anisotropic Expansion}
    A full six-dimensional search for anisotropic expansion is accurate, but is computational demanding. For the 
purpose of our study we also present three constrained versions of anisotropic expansion that may provide a more 
efficient search for the free energy minimum, and test to see how well they perform.

\subsubsection{3-Dimensional Anisotropic Expansion}
    The primary factors in thermal expansion is elongation of the lattice vectors or diagonal elements of the crystal
lattice tensor, $\boldsymbol{C}$. While the off-diagonals can change during thermal expansion, we noticed in our initial
work that the greatest change is in the diagonal elements. To study the importance of the off-diagonal elements in QHA, 
we examine anisotropic expansion where the off-diagonal elements remain fixed. That is, the gradient of thermal 
expansion, $\frac{\partial C_{i}}{\partial T}$, will only be solved for elements $i = 1, 2, 3$ of the crystal lattice 
tensor and eq.~\ref{equation:Gradient_Aniso} becomes an equation in a 3 dimensional vector of box lengths.

\subsubsection{1-Dimensional Anisotropic Expansion}
    We also present a third option for anisotropic expansion that mimics the fixed expansion in isotropic expansion, but
utilizes anisotropic thermal expansion ratios computed at 0 K. This approach is based on the same philosophy as the 
constant Gr\"{u}neisen parameter approach, which assumes that changes over most of the temperature range of interest 
are linear extrapolations of the behaviors at $T = $0 K, which anecdotally is a reasonable approximation. At the initial 
step of this variant of QHA, we determine the ratio in which values of $C_{i}$ change due to thermal expansion. We do 
this by solving eq.~\ref{equation:Gradient_Aniso} for the lattice minimum structure at 0 K to find  
$\kappa_i = \left(\frac{\partial C_{i}}{\partial T}\right)_{T = 0\:K}$, introducing the variable $\kappa_i$ as the derivative 
of $C_{i}$ with respect to $T$ at 0 K. Using the initial lattice structure, $\boldsymbol{C}(T = 0\:K)$, and the 
computed rate of thermal expansion, we can re-write $C_{i}$ as a function of a scaling factor $\lambda$, shown in 
eq.~\ref{equation:CM_lambda}.
  \begin{eqnarray}
    C_{i} (\lambda) = C_{i}(\lambda = 0) + \kappa_i \lambda(T) \label{equation:CM_lambda}
  \end{eqnarray}
Where we parameterize $\lambda$ such that $C_{i}(\lambda = 0) = C_{i}(T = 0\:K)$. More complicated parameterizations of 
$C_{i}(\lambda)$ can easily be defined, but we restrict this paper to the linear expansion given in 
eq.~\ref{equation:CM_lambda}. 

    Similarly to isotropic and anisotropic thermal expansion, we can determine the thermal gradient of $\lambda$ by
solving eq.~\ref{equation:dlambda_dT}.
  \begin{eqnarray}
    \frac{\partial \lambda}{\partial T} = \frac{\frac{\partial S}{\partial \lambda}}{\frac{\partial^{2} G}{\partial \lambda^{2}}}
    \label{equation:dlambda_dT}
  \end{eqnarray}
Eq.~\ref{equation:CM_lambda} is now a ordinary differential equation in $\lambda$. We can solve for $\lambda(T)$, which 
will give us the box vectors $C_i$ that minimize the free energy subject to the constraint given by 
Eq.\ref{equation:CM_lambda}.

\subsubsection{Quasi-Anisotropic Expansion}
    Our last anisotropic approach has been presented in literature in several similar 
forms~\cite{erba_combining_2014,cervinka_ab_2017} and provides an avenue to perform the stepwise thermal expansion 
approach while including some anisotropy within the crystal lattice. Similar to the stepwise approach, the crystal 
structure is compressed and expanded to several volume fractions around the lattice minimum structure; however, each 
compressed/expanded crystal structure is minimized at fixed volume find the minimum 0 K lattice geometry at that 
specified volume.

\section{Methods}
    We compare the efficiency and accuracy of eleven lattice dynamic calculations using combinations of the procedures 
above. Eight of these are new approaches based on the thermal gradient approach. These methods, and the abbreviations 
are shown in Table~\ref{table:methods_tested}.
  \begin{table*}[!htb]
  \begin{tabular}{ |l|P{6cm}|P{6cm}| }
    \hline
    \multicolumn{3}{ |c| } {Methods Tested} \\
    \hline
    Expansion &  Full Hessian & Gr\"{u}neisen or Polynomial Fit \\ \hline
    {None} & Harmonic approximation (HA) & \\ \hline
    {Isotropic} & Stepwise isotropic QHA (StepIso-QHA) & Stepwise isotropic QHA assisted with the Gr\"{u}neisen parameter (StepIso-QHA$\gamma$) \\ \cline{2-3}
                & Gradient isotropic QHA (GradIso-QHA) & Gradient isotropic QHA assisted with the Gr\"{u}neisen parameter (GradIso-QHA$\gamma$) \\ \hline
    {Non-Isotropic} & 1-Dimensional Gradient anisotropic QHA (1D-GradAniso-QHA) & 1-Dimensional Gradient anisotropic QHA assisted with the Gr\"{u}neisen parameter (1D-GradAniso-QHA$\gamma$) \\ \cline{2-3}
                    & 3-Dimensional Gradient anisotropic QHA (3D-GradAniso-QHA) & 3-Dimensional Gradient anisotropic QHA assisted with the Gr\"{u}neisen parameter (3D-GradAniso-QHA$\gamma$) \\ \cline{2-3}
                    & Unconstrained Gradient anisotropic QHA (GradAniso-QHA) & Unconstrained Gradient anisotropic QHA assisted with the Gr\"{u}neisen parameter (GradAniso-QHA$\gamma$) \\ \cline{2-3}
					& & Quasi-anisotropic QHA (QuasiAniso-QHA) \\ \hline
  \end{tabular}
  \caption{The eleven methods tested in this paper categorized by the type of thermal expansion and the approach by which Hessians are estimated.
  \label{table:methods_tested}}
  \end{table*}

\subsection{Crystal Polymorph Systems Tested} 
\label{section:Lattice_Structures}
    We selected resorcinol and piracetam to test all eleven methods on, both of which were studied in our previous 
work~\cite{dybeck_capturing_2017}. The two systems were chosen because StepIso-QHA determined that the lattice vectors only 
changed by 1--2\% from 0 to 300 K, but for MD, where anisotropy is considered, the lattice vectors change by 
-1--5\%. Additionally, it was found that piracetam form I had a box angle change with increasing temperature, which 
cannot be observed with an isotropic thermal expansion model.
  \begin{figure*}[!htb]
    \begin{center}
    \includegraphics[width=12cm]{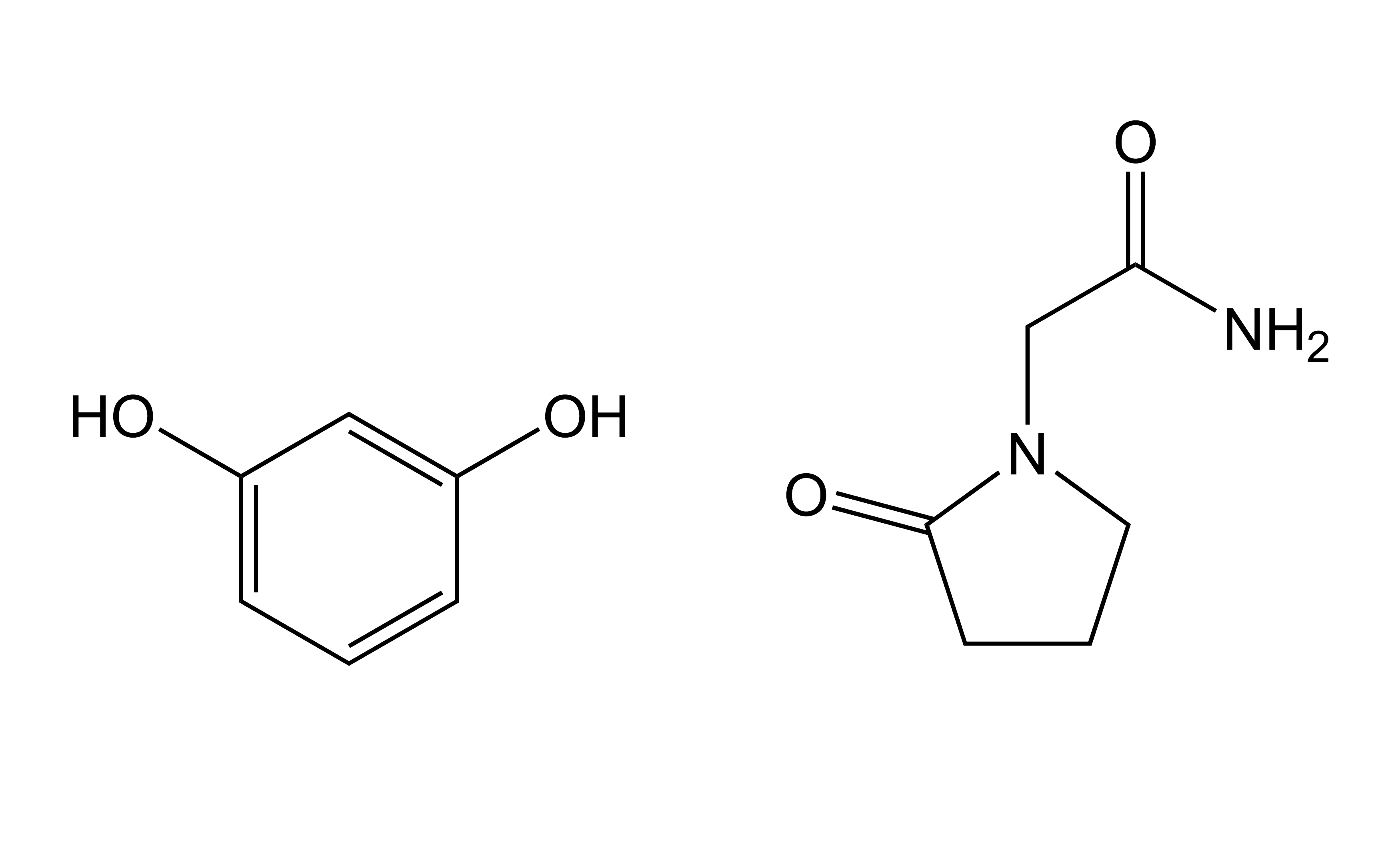}
    \end{center}
    \caption{Polymorphic systems evaluated in this paper using variants of QHA are resorcinol (left) and piracetam (right). 
    \label{figure:Molecules_Studied}}
  \end{figure*}

\begin{figure*}[!htb]
    \begin{center}
    \includegraphics[width=10cm]{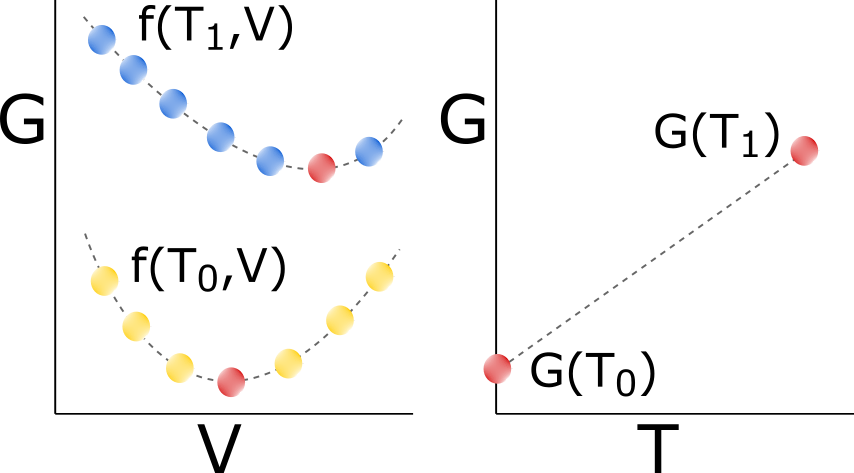}  
    \end{center}
    \caption{Schematic of stepwise isotropic QHA. The $G$ vs.~$V$ plot (left) shows $f(V,T)$ at two temperatures, $T_{0}$ 
             and $T_{1}$. At $T_{0}$ the lowest energy structure, in red, is isotropically expanded and compressed, with 
             each structures free energy computed $f(T_{0},V)$ is formed. The free energy of each structure is 
             re-evaluated at $T_{1}$, generating $f(T_{1},V)$. At each temperature $i$, the lowest free energy 
             structure is chosen for each $i$ by minimizing $f(T_{i},V)$, and the free energy at $T_i$ is set to this minimum, 
			 allowing the construction of $G(T)$ shown (right).
    \label{figure:stepwise_QHA}}
  \end{figure*}

\textit{Stepwise Isotropic QHA (StepIso-QHA)}: \label{section:StepIso-QHA} For stepwise isotropic QHA, 
  \begin{enumerate}
    \item the initial lattice minimum structure is expanded isotropically by a small volume fraction;
    \item the new expanded system is geometry optimized;
    \item steps 1 and 2 are repeated on the expanded structure; and
    \item the process is continued until the crystal is expanded past the anticipated volume at the maximum temperature.
  \end{enumerate}
Computing the lattice energy of all expanded structures produces a lattice energy curve $U(V)$. The mass-weighted 
Hessian of each expanded structure is computed and diagonalized to provide $\omega_{k} (V_{i})$, allowing the 
computation of a vibrational surface $A_{v}(V,T)$. At each temperature the lattice volume, that minimizes the Gibbs 
free energy is used to solve for $G(T)$ in eq.~\ref{equation:QHA_G_iso}. A schematic of the stepwise approach is 
shown in fig.~\ref{figure:stepwise_QHA}. 
    For our work, we chose to compress and expand each crystal structure to volume fractions ($(V + \Delta V) / V$) of 
0.99 and 1.08 respectively, which is sufficiently large for the crystals examined here. Intermediate volumes
between 0.99 and 1.08 were evaluated at values of $\Delta V / V = 10^{-3}$. We discuss how the choice of $\Delta V / V$
effects the smoothness of $V(T)$ and $G(T)$ in the Supporting Information (Section~\ref{section:StepIso_Spacing}).

    While the result in fig.~\ref{figure:stepwise_QHA} is straightforward to calculate for isotropic expansion, it 
becomes considerably more expensive for anisotropic expansion. The one dimensional problem of isotropic expansion 
scales with the number, $N$, of sampled points, $\{V_{0}, V_{1}, ..., V_{N}\}$. For anisotropic expansion, the problem 
is now six dimensional, which will scale as $N^{6}$ in order to determine the minimum free energy structure for all six 
lattice parameters. While not every sample is required, as some can be ruled out as being too far from the minimum, we 
in general need some understanding of how the crystal expands before limiting the $N^{6}$ dimensional space to 
something more feasible.

\textit{Stepwise Isotropic QHA with the Gr\"{u}neisen Parameter (StepIso-QHA$\gamma$)}: \label{section:StepIso-QHA+Gru}
    Similar to StepIso-QHA, stepwise isotropic QHA with the Gr\"{u}neisen parameter involves 
isotropic expansion of  the initial lattice minimum structure by a small steps to produce the same lattice energy curve 
in StepIso-QHA. Where StepIso-QHA and StepIso-QHA$\gamma$ differ is the calculation of the vibrational spectra of every 
structure. StepIsoQHA$\gamma$ circumvents multiple Hessian calculations by approximating changes in the vibrational 
spectra due to expansion by assuming the Gr\"{u}neisen parameter, calculated using eq.~\ref{equation:Gru_Iso_solution}, 
is independent of box volume, meaning it can be calculated using eq.~\ref{equation:Gru_Iso_new}. This significantly
decreases the computational cost compared to StepIso-QHA. 

\begin{figure*}[!htb]
    \begin{center}
    \includegraphics[width=10cm]{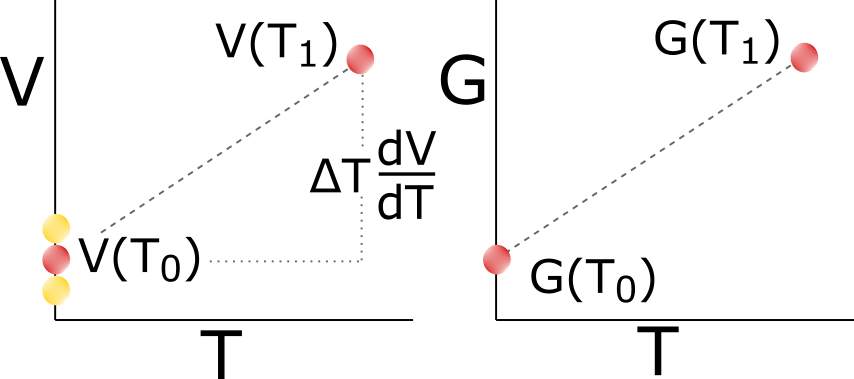}
    \end{center}
    \caption{Schematic of gradient isotropic QHA. On the $V$ vs.~$T$ plot (left) the known lattice minimum structure with 
             volume $V(T_{0})$ is shown along the y-axis as the red dot. To determine the gradient of expansion using 
             finite differences, the structure is expanded and compressed, indicated by two yellow dots on the y-axis, and 
			 $\partial S / \partial V$ and $\partial^2 G / \partial V^2$ are estimated. With each structure's Gibbs free 
			 energy and entropy, we can determine $\frac{\partial V}{\partial T}$ using eq.~\ref{equation:Gradient_Iso}, and 
			 with a given temperature step, $\Delta T$, the volume of the 
             structure at $T_{1}$ can be computed with numerical integration. The structure at $T_{0}$ can be directly 
             expanded to $V(T_{1})$ and the free energy of the expanded structure can be computed to compute $G(T_1)$ 
             (right).
    \label{figure:gradient_QHA}}
  \end{figure*}

\textit{Gradient Isotropic QHA (GradIso-QHA)}: \label{section:GradIso-QHA}
    Gradient QHA provides a more efficient way to calculate free energy curves and the gradient method can be easily 
performed for anisotropic expansion. Fig.~\ref{figure:gradient_QHA} depicts how gradient isotropic QHA works. At 0 K 
the gradient of thermal expansion is determined by isotropically expanding and compressing the lattice minimum 
structure to solve eq.~\ref{equation:Gradient_Iso}. 

    Comparing figures \ref{figure:gradient_QHA} and \ref{figure:stepwise_QHA} shows that a gradient approach requires only
the structures local to the minimum at any given temperature, while the stepwise approach can require many more. We also 
note that the stepwise approach only produces a smooth $V(T)$ curve if there is minimal spacing between expanded 
structures, while the gradient approach only relies on the numerical method used. Since we do not have an analytical 
solution to eq.~\ref{equation:Gradient_Iso}, we will need to solve for $\frac{\partial V}{\partial T}$ numerically. We 
implement a 4$^{th}$ order Runge-Kutta method to accurately determine the free energy curve across the entire 
temperature range.

    The solution of $\frac{\partial V}{\partial T}$ is solved by computing $\frac{\partial S}{\partial V}$ and 
$\frac{\partial^{2} G}{\partial V^{2}}$ with a central finite difference approach. The lattice minimum structure is
isotropically expanded and compressed, the mass-weighted Hessian is computed and diagonalized for the vibrational 
spectra of all three structures, and the Gibbs free energy is evaluated to determine the numerical gradient. For 
GradIso-QHA we found that only three Runge-Kutta steps of 100 K are required to produce a visually smooth volume versus 
temperature curve up to 300 K. To solve for intermediate temperature points, we used a 3$^{rd}$ order spline on the volume 
parameter, which uses the first Runge-Kutta gradient at $T_{i}$ and $T_{i+1}$ to determine $V(T)$ for 
$T_{i} \leq T \leq T_{i+1}$.~\cite{hairer_solving_2008} Further information on step sizes can be found in the 
Supporting Information (Section~\ref{section:Euler_RK}).

\textit{Gradient Isotropic QHA with the Gr\"{u}neisen Parameter (GradIso-QHA$\gamma$)}: \label{section:GiQg}
    Gradient isotropic QHA with the Gr\"{u}neisen Parameter uses the same local gradient approach
as GradIso-QHA, with replacement of the calculation of the vibrational frequencies through the Hessian with the calculation using the 
Gr\"{u}neisen parameter approximation. Before any gradients are calculated, the Gr\"{u}neisen parameters are calculated 
using eq.~\ref{equation:Gru_Iso_solution}. The local gradient is calculated the same way as in GradIso-QHA, except that 
instead of calculating the mass-weighted Hessian of the isotropically expanded structures, 
eq.~\ref{equation:Gru_Iso_new} provides the new vibrational spectra.

\textit{Quasi-anisotropic QHA (QuasiAniso-QHA)}: \label{section:QuasiAniso-QHA}
    Quasi-anisotropic QHA uses a similar methodology to that implemented in the CRYSTALS program~\cite{crystal_manual} 
as well variants used by other researchers~\cite{cervinka_ab_2017}. The lattice minimum structure is first isotropically 
expanded and compressed to several volumes of interest. We then perform a volume-constrained optimization of the lattice 
energy for each volume, and then fit the potential energy with an equation of state and a polynomial fit to each 
vibrational frequency. We then use the fitted functions of the energy and vibrational modes to minimize the Gibbs free 
energy at each temperature of interest.

    For our work, we chose to compress and expand each crystal structure to volume fractions ($(V + \Delta V) / V$) of 
0.99 and 1.05 respectively. Intermediate volumes between 0.99 and 1.05 were evaluated at values of  
$\Delta V / V = 10^{-2}$, which is an order of magnitude larger than StepIso-QHA since we are using a polynomial fit. 
We then perform a volume-constrained optimization of the lattice energy and fit the $U(V)$ with the Rose-Vinet equation 
of state. This optimization was implemented as a bounded constrained optimization in Python wrapped around TINKER energy 
call, more information can be found in the Supporting Information (Section~\ref{section:constrained_optimization}). 
This is clearly not the fastest way to perform the optimization, but allowed us to perform the optimization without 
changing the underlying code base. The vibrational frequencies of all structures are computed and each frequency is 
fitted with a 3$^{rd}$ order polynomial. We then use these these functions to determine the volume that minimizes the 
Gibbs free energy, eq.~\ref{equation:QHA_G_aniso}. We verified that constrained volume optimizations were at the 
correct minimum by using the method of Lagrange multipliers.

\textit{Gradient Anisotropic QHA (GradAniso-QHA)}: \label{section:GradAniso-QHA}
    Modeling anisotropic expansion includes a greater number of variables to minimize along, making it computationally  
costly compared to isotropic expansion. We have therefore chosen to apply only the gradient method to anisotropic QHA 
because the stepwise approach, which would require enumeration of the points on a 6D grid would be too expensive. 
Gradient anisotropic QHA uses a similar approach as GradIso-QHA, except instead of determining the 
volumetric thermal gradient, we compute the gradient of the six lattice parameters. We solve 
$\frac{\partial \boldsymbol{C}}{\partial T}$ in eq.~\ref{equation:Gradient_Aniso} using a finite difference approach 
and apply a 4$^{th}$ order Runge-Kutta method to converge onto the structure that minimizes the Gibbs Free energy.

    $\frac{\partial \boldsymbol{C}}{\partial T}$ is computed from the solution of $\frac{\partial S}{\partial C_{i}}$ 
and $\frac{\partial^{2} G}{\partial C_{i}^{2}}$ with a central finite difference approach. For each lattice parameter, 
the lattice minimum structure is expanded or compressed ($C_{i} \pm dC_{i}$) while holding the other lattice parameters 
constant. The altered structures are geometry optimized, the mass-weighted Hessian is computed and diagonalized for the 
vibrational spectra for all structures, and the Gibbs free energy is evaluated to determine the numerical gradient. In the 
case where the lattice minimum structure has a value of $C_{i} = 0$, we set $dC_{i} =0$. Similar to GradIso-QHA, we 
found that three Runge-Kutta steps produced a structure at a free energy minimum for all temperature and intermediate steps 
were solved with a 3$^{rd}$ order spline. For further information regarding the choice of step sizes used in the finite 
difference, refer to the Supporting Information (Section~\ref{section:Euler_RK}).

\textit{Gradient Anisotropic QHA with the Gr\"{u}neisen Parameter (GradAniso-QHA$\gamma$)}: \label{section:GradAniso-QHA+Gru}
    Gradient anisotropic QHA with the Gr\"{u}neisen Parameter uses the same local gradient 
approach as GradAniso-QHA with the replacement of the Gr\"{u}neisen parameter to approximate the vibrational spectra of
all expanded structures.

\textit{3D-Gradient Anisotropic QHA (3D-GradAniso-QHA)}:
    Three dimensional gradient anisotropic QHA follows the same approach as GradAniso-QHA, except 
components $i = 4, 5, 6$ for $\frac{\partial C_{i}}{\partial T}$ are zero. This means that in eq.~\ref{equation:Gradient_Aniso}, 
$\frac{\partial \boldsymbol{C}}{\partial T}$ and $\frac{\partial S}{\partial \boldsymbol{C}}$ are three component 
vectors and $\frac{\partial^{2} G}{\partial \boldsymbol{C}^{2}}$ is reduced to a 3$\times$3 matrix. We then integrate
up with temperature by taking Runge-Kutta step sizes of 100 K up to 300 K and determine intermediate points with a 3$^{rd}$ 
order spline, which is the same as what was done in isotropic and unconstrained anisotropic expansion.

\textit{3D-Gradient Anisotropic QHA with the Gr\"{u}neisen Parameter (3D-GradAniso-QHA$\gamma$)}:
    Three dimensional gradient anisotropic QHA with the Gr\"{u}neisen Parameter uses the same 
local-gradient approach as 3D-GradAniso-QHA with the addition of the Gr\"{u}neisen parameter to approximate the 
vibrational spectra of all expanded structures. 

\textit{1D-Gradient Anisotropic QHA (1D-GradAniso-QHA)}:
    One dimensional gradient anisotropic QHA allows us to capture anisotropic expansion, with little
additional computational cost compared to GradIso-QHA. With the lattice minimum structure, we compute 
$\frac{\partial \boldsymbol{C}}{\partial T}$ at 0 K, which provides a fixed ratio $\kappa_{i}$ for how the lattice
parameters can change relative to each other with temperature. In eq.~\ref{equation:CM_lambda}, 
$C_i(\lambda = 0)$ at 0 K and $\frac{\partial \lambda}{\partial T}$ can be computed to determine 
how $\lambda$ changes with temperature.

    We first compute $\left(\frac{\partial C_{i}}{\partial T}\right)_T = 0\:K$ in the same way performed in GradAniso-QHA,
but this is performed only on the lattice minimum structure to get the values of $\kappa_{i}$. The solution of 
$\frac{\partial \lambda}{\partial T}$ is solved by computing $\frac{\partial S}{\partial \lambda}$ and 
$\frac{\partial^{2} G}{\partial \lambda^{2}}$ with a central finite difference approach. The lattice minimum structure 
is expanded and compressed by changing $\lambda$ in eq.~\ref{equation:CM_lambda} by $\pm d\lambda$, the mass-weighted 
Hessian is computed and diagonalized for the vibrational spectra of all three structures, and the Gibbs free energy is 
evaluated to determine the numerical gradient. We then integrate up with temperature by taking Runge-Kutta step sizes of 
100 K up to 300 K and determine intermediate points with a 3$^{rd}$ order spline, which is the same as is done in 
isotropic and unconstrained anisotropic expansion.

\textit{1D-Gradient Anisotropic QHA with the Gr\"{u}neisen Parameter (1D-GradAniso-QHA$\gamma$)}:
    One dimensional gradient anisotropic QHA with the Gr\"{u}neisen Parameter (1D-GradAniso-QHA$\gamma$) uses the same 
local-gradient approach as 1D-GradAniso-QHA with the addition of the Gr\"{u}neisen parameter to approximate the 
vibrational spectra of all expanded structures. 

\subsection*{Simulation Details} \label{section:Simulation_Details}
    All calculations were performed using our Python based lattice dynamics code available on GitHub at 
\url{http://github.com/shirtsgroup/Lattice_dynamics}. The code is currently built to run for a test systems and the Tinker 
8.1 molecular modeling package~\cite{ponder_efficient_1987} for vibrational spectra, lattice minimizations, and 
potential energy calculations. The code is designed modularly to be adapted to any program that performs energy 
minimizations, potential energy and vibrational outputs; we are in the process of incorporating Quantum ESPRESSO.~\cite{giannozzi_quantum_2009} 

    Lattice structures were retrieved from the Cambridge Crystallographic Data Center 
with Tinker's {\tt xtalmin} executable to an RMS gradient/atom of $10^{-5}$. When the crystals are expanded, the center 
of mass for each molecule is moved, preserving the intramolecular distances. The expanded structures are geometry 
optimized using Tinker's {\tt minimize} executable with an RMS gradient/atom of $10^{-4}$ to geometry optimize the 
crystal. Discussion of minimization criterion can be found in the Supporting Information. In some cases Tinker aborted 
before this point; both cases for Tinker aborting and choice of RMS is discussed further in the Supporting Information
(Section~\ref{section:Tink_ViqFreq}). 

    For each crystal structure, the lattice energy ($U$) is computed with Tinker's {\tt analyze} executable. In the cases 
where the mass-weighted Hessian is calculated, we compute the vibrational spectra using Tinker's {\tt vibrate} 
executable. We have edited Tinker's code to perform a finite difference calculation to compute the second derivative of 
the charge potential energy function. Discussion of of our code adjustments can be found in the Supporting Information
(Section~\ref{section:Mod_Tinker}).

    The isotropic Gr\"{u}neisen parameters were found by solving eq.~\ref{equation:Gru_Iso_solution} using the 
{\tt vibrate} executable on the lattice minimum structure and the lattice minimum structure expanded by 
$\Delta V / V = 1.5 \times 10^{-3}$. The vibrational spectra of all other isotropically expanded structures were found 
using eq.~\ref{equation:Gru_Iso_new}. The anisotropic Gr\"{u}neisen parameters were found by solving 
eq.~\ref{equation:Gru_Iso_solution} in all six of the strain directions. For each strain, the crystal matrix parameter 
($\eta_{i}$; $i = 1, 2, ... 6$) is increased by $\eta_{i} = 1.5\times10^{-3}$ and using the {\tt vibrate} executable the 
vibrational spectra of the lattice minimum structure and 6 strained structures are calculated. The vibrational spectra 
of all other strained structures are found using eq.~\ref{equation:Gru_Aniso_new_strain_full}. We also assure that
the vibrational modes match correctly between differing volumes, further information can be found in the Supporting
Information (Section~\ref{section:matching_modes}).

    Determining the correct numerical step size for computing the gradient of thermal expansion is the main factor in
assuring that the crystal remains at a Gibbs free energy minimum up to the temperature of interest. When computing the
thermal gradient ($\frac{\partial y}{\partial T}$ where $y = V$, $\lambda$, or $C_{i}$) there is some finite step size
($\pm \Delta y$) that is required to solve the thermal gradient numerically. We found that the best way to determine $\Delta y$ for
each crystal and method was to pick a step size that would change the lattice minimum energy by $5\times10^{-4}$ kcal/mol. 
Further discussion of this and a table of the values can be found in the Supporting Information 
(Section~\ref{section:Gradi_step_size}).

\section{Results and Discussion}
\subsection{Converging on the Minimum Gibbs Free Energy Structure}
    The quasi-harmonic approximation computes the Gibbs free energy at a given temperature by minimizing $y$ in the 
function $f(y,T)$. In this approach, $y$ can be the isotropic volume ($V$), the crystal lattice tensor 
($\boldsymbol{C}$), or the scalar $\lambda$ used in our 1D-anisotropic approach. With the gradient approach, we can 
easily determine if the structure that is computed to minimize $f$ is at a local minimum by estimating if 
$\frac{\partial G}{\partial y} = G^{\prime} = 0$. Our code provides an approximate check of whether a point is nearly 
at the minimum by comparing the backwards, central, and forward finite solutions of $\frac{\partial G}{\partial y}$.
Since the first-order errors in $G^{\prime}_{backwards}$ and $G^{\prime}_{forwards}$ should be equal and opposite, 
if the ordering $G^{\prime}_{backwards} < G^{\prime}_{central} < G^{\prime}_{forwards}$ is satisfied then at least at that 
level of error we are at the minimum.

    Most of the gradient approaches described above converged to minimum free energy structures within numerical error
at every Runge-Kutta step for all four crystal structures. The only methods that failed to do this for some crystal 
polymorphs were GradIso-QHA, GradAniso-QHA$\gamma$, and 1D-GradAniso-QHA for piracetam form III as well as 
1D-GradAniso-QHA for resorcinol form $\alpha$. For GradIso-QHA with piracetam form III and 1D-GradAniso-QHA with 
resorcinol form $\alpha$, we were able to achieve our minimization criterion at every step by increasing the number of 
Runge-Kutta steps up to 300 K. For both of these methods on these systems only, the results in the following sections 
will be those achieved with four Runge-Kutta steps instead of three. For the other two methods performed on piracetam 
form III, we found that the crystal structures were re-structuring at $T > 0\:K$, indicating a failure in the QHA 
approach.

\subsection{Discontinuities in the Free Energy Surface are Problematic for QHA} \label{section:results_discontinuities}
    QHA will fail when there are multiple minima on the free energy surface which can interconvert when lattice 
expansion occurs. QHA assumes that there is a single minimum that represents the thermodynamic properties at a given 
temperature and pressure, which becomes problematic when there are multiple minima present. In our previous work with 
MD simulations, we observed that a number of crystal structures re-minimized to new structures when the crystal was 
heated.~\cite{dybeck_comparison_2016} However, that is not what is generally happening here. Instead, we have found that in the 
process of expansion, even relatively small restructuring can lead to a failure of both the structure and free energy 
to vary continuously, thus violating the assumptions of QHA.
  \begin{figure*}[!htb]
    \begin{center}
    \includegraphics[width=16cm]{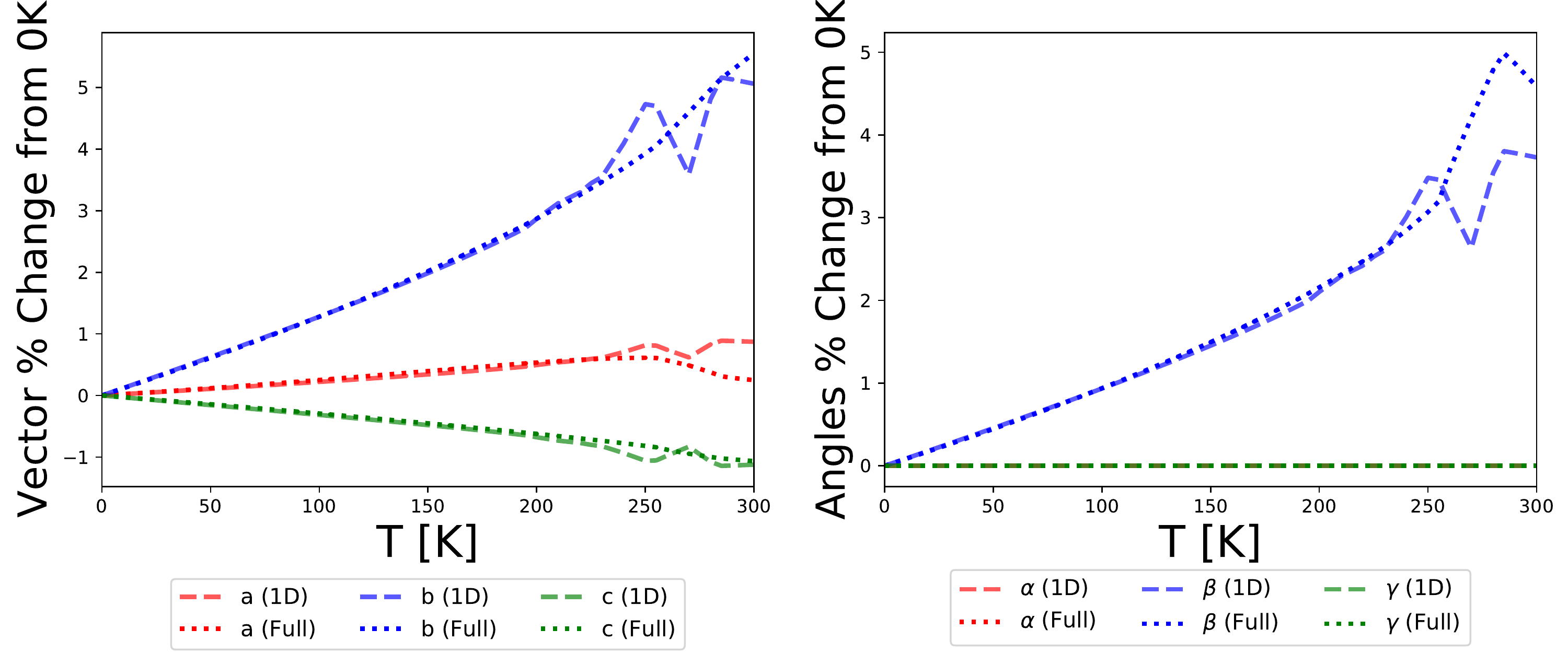} \\
    \includegraphics[width=16cm]{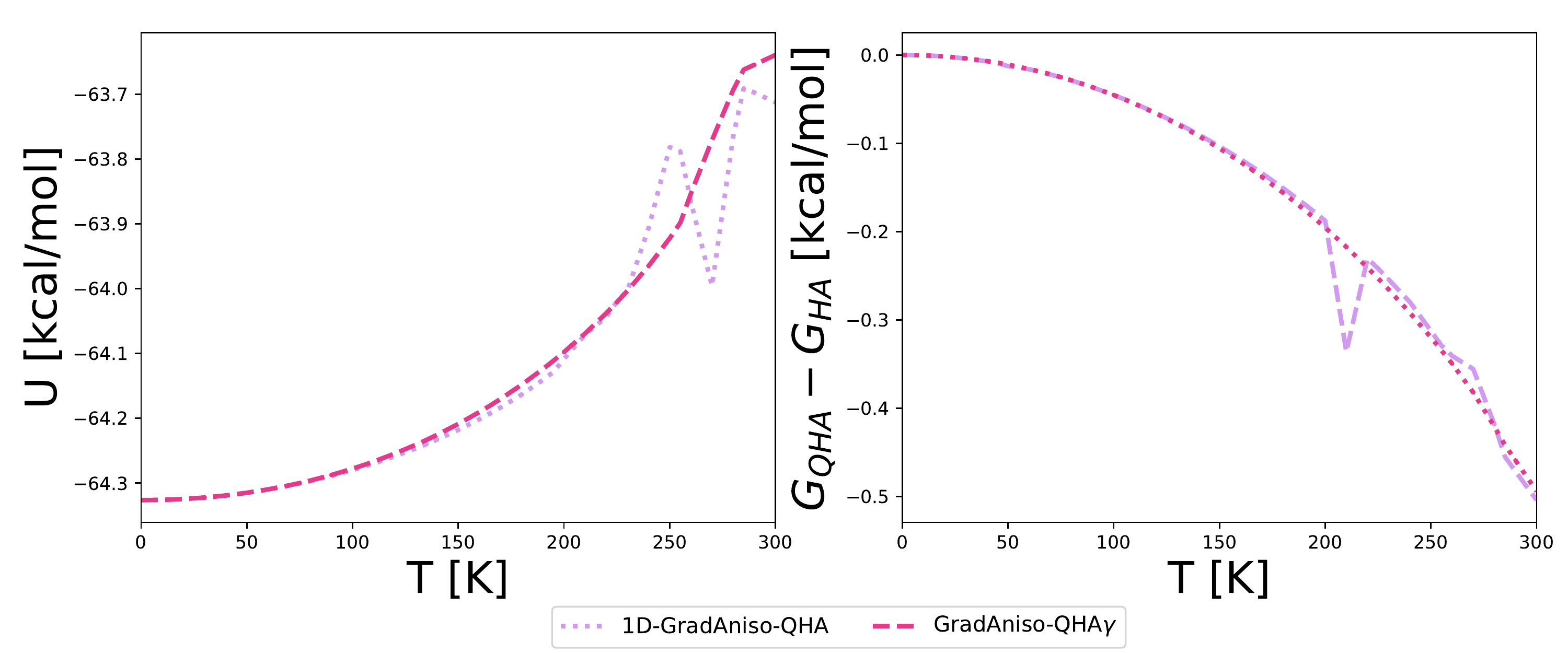}
    \end{center}
    \caption{Computed percent change of lattice parameters (top left and right) from 0 K using 
             eq.~\ref{equation:lattice_expansion}, potential energy $U$ (lower left), and $G_{QHA} - G_{HA}$ (lower 
             right) for piracetam form III using 1D-GradAniso-QHA (1D) and GradAniso-QHA$\gamma$ (Full). Both methods 
             produce similar, smooth results up to 200 K. After 200 K, the crystal re-structures for both methods, and 
             additionally falls out outside of the free energy minimum as determined by numerical checks of 
             $\partial G / \partial C_{i}$. 
    \label{figure:discont}}
  \end{figure*}
    We looked at how the potential energy, free energy, and lattice vectors and angles change with temperature to help 
quantify the existence of a discontinuity found from thermal expansion. For piracetam form III, we took 20 Runge-Kutta 
steps up to 300 K for 1D-GradAniso-QHA and GradAniso-QHA$\gamma$ to help pinpoint where the QHA was failing for both
methods. The top two graphs in fig.~\ref{figure:discont} present the percent change in the lattice parameters, 
$\%h$, for the three lattice vectors (left) and angles (right). For a particular lattice parameter, $h_{i}$, the percent 
change is calculated as:
  \begin{eqnarray}
   \% h_{i}(T) = \frac{h_{i}(T) - h_{i}(T=0\:K)}{h_{i} (T=0\:K)}\times100\%
    \label{equation:lattice_expansion}
  \end{eqnarray}
The bottom two plots in fig.~\ref{figure:discont} show the potential energy ($U$) and the Gibbs free energy 
deviation from HA ($G_{QHA} - G_{HA}$). Up until 200 K, both methods provide visually indistinguishable and smooth
results for the changes in lattice parameters and energy with temperature. Additionally, our methods produce 
structures at a free energy minimum ($\frac{\partial G}{\partial \lambda} = 0$ or 
$\frac{\partial G}{\partial \boldsymbol{C}} = 0$) up to 255 K for 1D-GradAniso-QHA and to 270 K for 
GradAniso-QHA$\gamma$. After these temperatures the computed lattice parameters and energies have rough transitions 
with temperature and oscillate.

\begin{figure*}[!htb]
    \begin{center}
    \includegraphics[width=8cm]{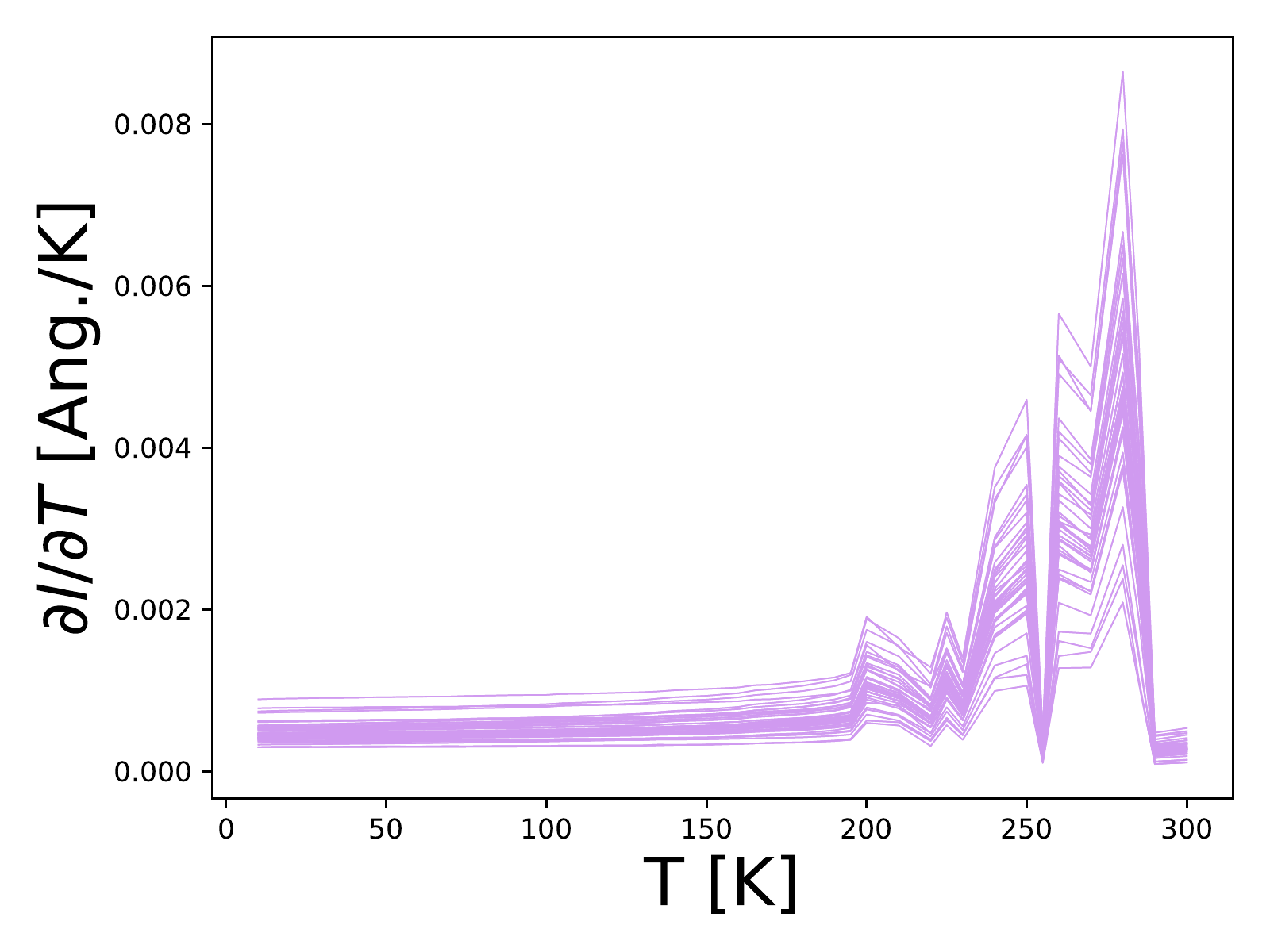}
    \includegraphics[width=8cm]{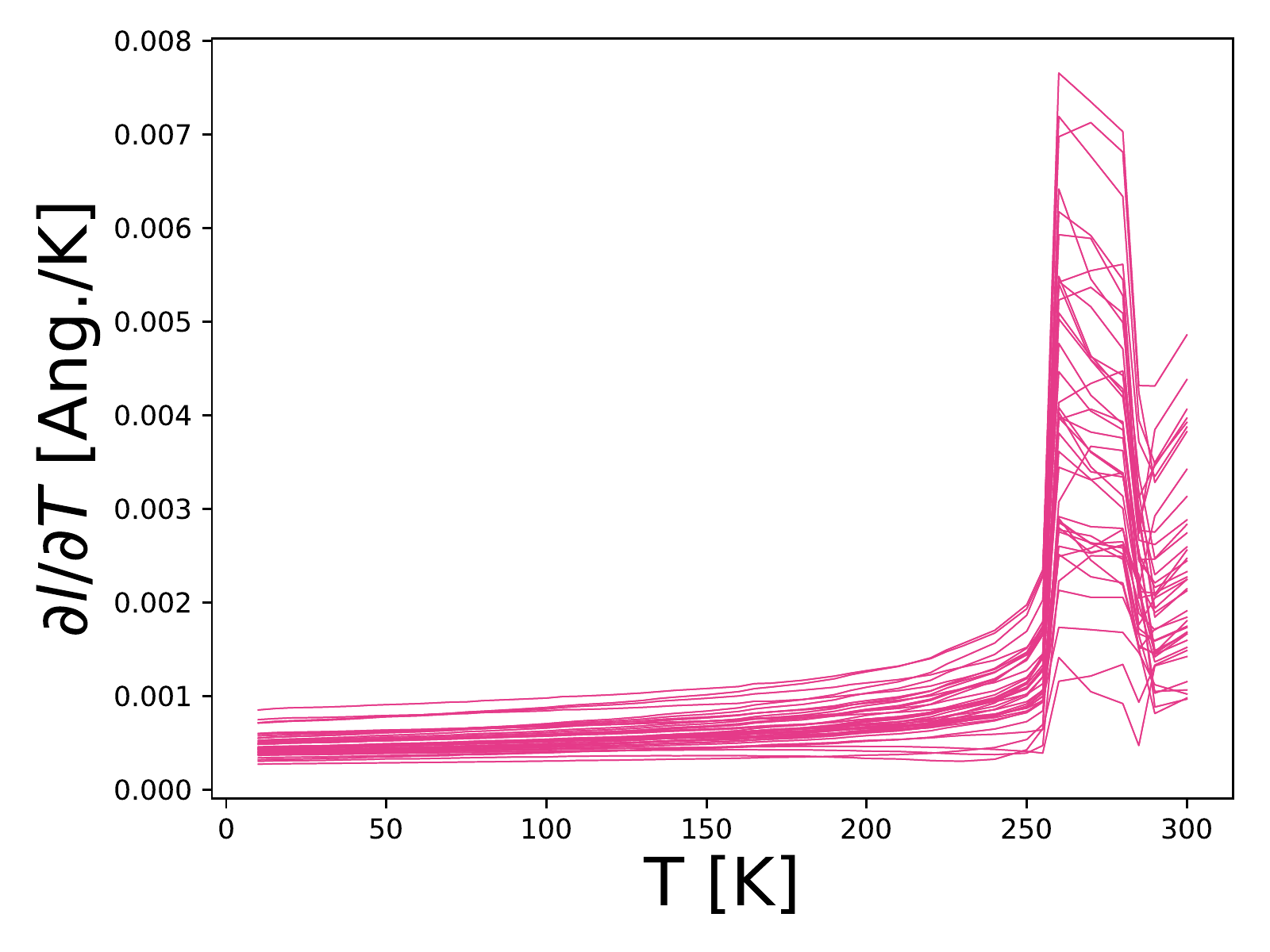}
    \end{center}
    \caption{Computed $\frac{\partial l}{\partial T}$ for each atom in the crystal of piracetam form III using 
             1D-GradAniso-QHA (left) and GradAniso-QHA$\gamma$ (right) using eq.~\ref{equation:atom_velocity}. Each line
             represents a different atom within the unit cell. Up until that 200 K $\frac{\partial l}{\partial T}$ is 
             smooth, but after 200 K the behavior becomes rough and sharply increases implying a re-structuring of the
             crystal. We use the same colors for each method that are used in the energy plots of 
             fig.~\ref{figure:discont}.
    \label{figure:discont_cord}}
  \end{figure*}

    Discontinuities in the free energy surface are most easily visualized by looking at each atoms movement due to 
thermal expansion as a function of $T$.  In Cartesian coordinates, a particular atom will have a position vector, 
$v = [x, y, z]$, that we can track with its temperature change by computing $\frac{\partial l}{\partial T}$ 
by backwards finite difference, which is shown in eq.~\ref{equation:atom_velocity}.
  \begin{eqnarray}
    \frac{\partial l}{\partial T} = \frac{\norm{v_{i} - v_{i-1}}}{T_{i} - T_{i-1}}
    \label{equation:atom_velocity}
  \end{eqnarray}
The change in $\frac{\partial l}{\partial T}$ as a function of temperature is shown in fig.~\ref{figure:discont_cord}.
Each line represents the motion of a different atom.

    The  sudden oscillation in the lattice parameters and energy are due to adjustments in the molecular arrangement 
within the crystal. As the crystal expands, new volumes open, which provides new low energy arrangements of individual 
atoms or entire molecules when geometry optimized. Fig.~\ref{figure:discont_cord} shows that this behavior is not 
smooth for some crystals. Up until 200 K 1D-GradAniso-QHA shows that $\frac{\partial l}{\partial T}$ changes smoothly 
and relatively slowly with temperature (left in fig.~\ref{figure:discont_cord}). After 200 K all of the atoms start to 
move dramatically faster, leading to the oscillation of lattice parameters and energy at 250 K. Similarly, GradAniso-QHA 
has a large shift in the coordinates at 250 K (right in fig.~\ref{figure:discont_cord}), which also coincides with 
roughness of the curves for the lattice parameters and energy. 

    For the free energy and lattice expansion of individual polymorphs we compare the results at 300 K, we therefore
do not show results for GradAniso-QHA, GradAniso-QHA$\gamma$, and 1D-GradAniso-QHA because the system becomes 
discontinuous before we reach 300 K. Since we show the polymorph free energy differences across the temperature range,
we do show GradAniso-QHA$\gamma$ and 1D-GradAniso-QHA up to 200 K to understand the low temperature behavior.

\subsection{Polymorph Gibbs Free Energy}
\begin{figure*}[!htb]
    \begin{center}
    \includegraphics[width=16cm]{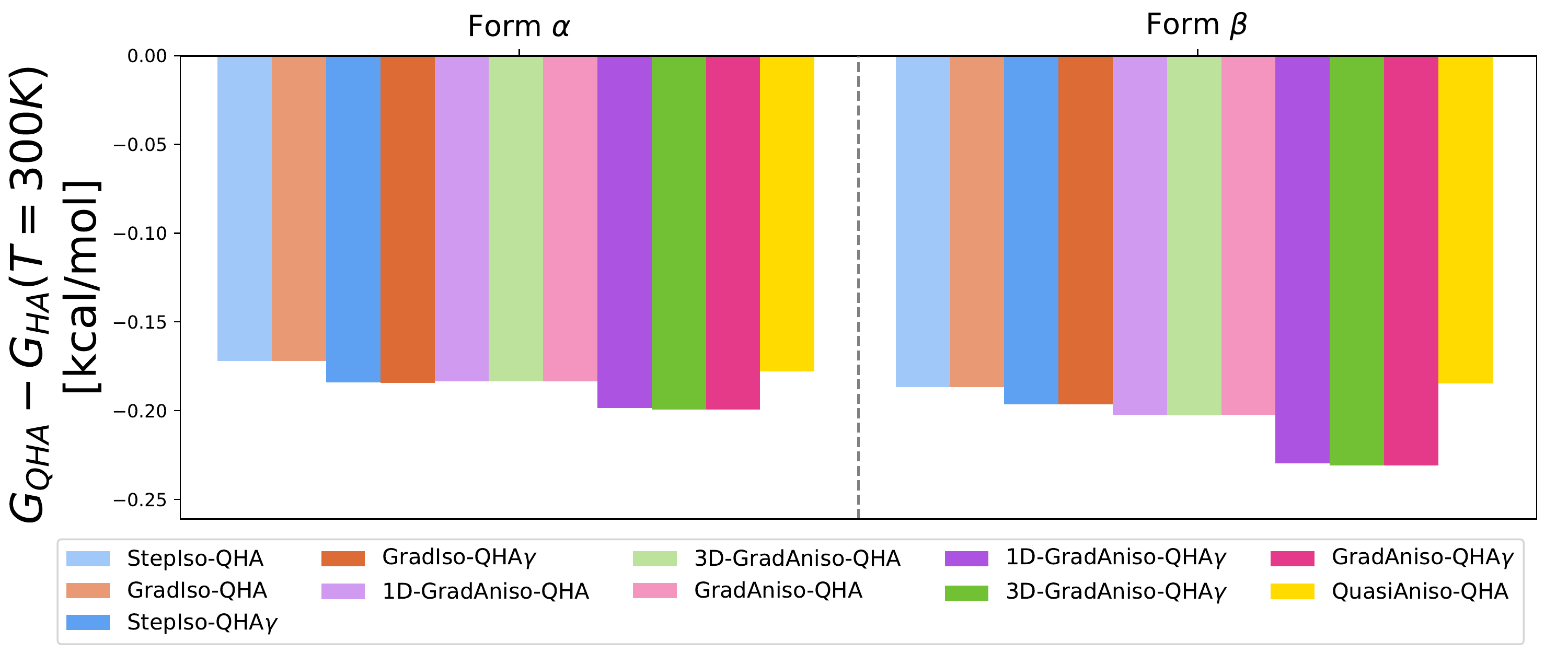} \\
    \includegraphics[width=16cm]{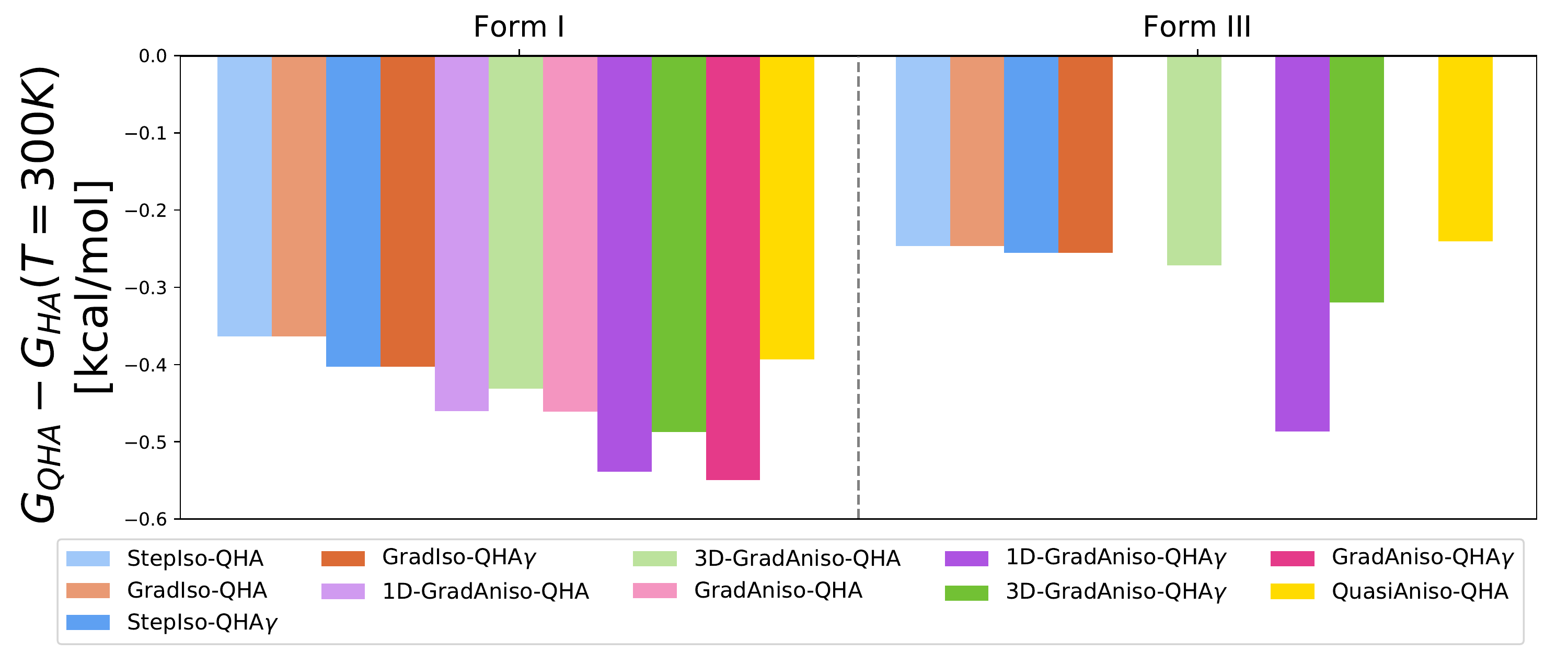}
    \end{center}
    \caption{The Gibbs free energy deviations of the QHA methods from HA which completed are shown at 300 K for resorcinol 
             (top) and piracetam (bottom). By removing constraints from the thermal expansion model, the crystal relax 
             into a lower free energy structure. The largest differences in free energy for the 10 QHA methods are due
             to using the Gr\"{u}neisen parameter. The results for 1D-GradAniso-QHA$\gamma$, GradAniso-QHA, and 
             GradAniso-QHA$\gamma$ for piracetam form III are not shown because of discontinuities in the free energy
             as a function of $T$, as discussed in section~\ref{section:results_discontinuities}.
    \label{figure:Gfinal}}
  \end{figure*}

    All results for the ten QHA methods are shown as their Gibbs free energy deviation from the harmonic approximation. 
The results for the full plots of $G$ versus temperature can be found in the Supporting Information 
(Section~\ref{section:polymorph_free_energies}). Instead, we will summarize the results at 300 K, as shown in fig.~\ref{figure:Gfinal}. As the 
deviations increase with temperature, the 300 K results represent the largest change.

    Our gradient method successfully matches the stepwise approach for isotropic expansion within reasonable numerical 
error. Although the gradient and stepwise approaches determine the minimum Gibbs free energy volume in different ways, 
both methods should produce similar results as they solve the same equation. For all four crystal structures the 
deviations between the stepwise and gradient approach are less than 0.0002 kcal/mol at 300 K, which is smaller than 
the required energy tolerances in literature for organic polymorphs.

    The Gr\"{u}neisen approaches have small, but noticeable effects on the predicted Gibbs free energy for isotropic 
expansion. For all four crystals, there is less than a 0.04 kcal/mol deviation between all isotropic methods.
This deviation is largest, 0.039 kcal/mol, for piracetam form I when comparing the computed free energy at 300 K for 
GradIso-QHA and GradIso-QHA$\gamma$. For all other crystals, the difference between Gr\"{u}neisen and non-Gr\"{u}neisen
approaches is $<$ 0.013 kcal/mol. 

    By removing constraints on the crystal lattice parameters and thus increasing the accuracy of the thermal expansion
we find that the crystal is able to expand into a lower Gibbs free energy minimum for all crystals, as would be 
anticipated as constraints are removed from a minimization. As a validation, the following ranking of free energy for 
methods should hold true for all $T$.
  \begin{eqnarray}
    G_{HA} > G_{Isotropic-QHA} \geq G_{Anisotropic-QHA}
  \end{eqnarray}
We see that the free energies in fig.~\ref{figure:Gfinal} indeed satisfy this condition for all crystals.

    For resorcinol, our three anisotropic methods converge to the same result within error, but the Gr\"{u}neisen 
approaches overestimate the decrease in free energy due to anisotropic thermal expansion. For the three different 
anisotropic methods there is a difference in the computed free energy that is less than 0.002 kcal/mol at 300 K. In the 
top graph of fig~\ref{figure:Gfinal} the only visible differences between the 6 anisotropic methods are between the 
three with full Hessian calculations and the Gr\"{u}neisen approaches (i.e. comparing 1D-GradAniso-QHA with 
1D-GradAniso-QHA$\gamma$, 3D-GradAniso-QHA with 3D-GradAnsio-QHA$\gamma$, and GradAnsio-QHA with 
GradAniso-QHA$\gamma$), all of which are $<$ 0.03 kcal/mol. 

    Piracetam has greater variability between our six anisotropic QHA methods, even among those that do not use the 
Gr\"{u}neisen parameter approximation. In the bottom graph of fig.~\ref{figure:Gfinal} the results for 1D-GradAniso-, 
GradAniso-QHA, and GradAniso-QHA$\gamma$ are not shown for form III because of the discontinuities discussed in section 
\ref{section:results_discontinuities}. The 1D-, 3D-, and unconstrained anisotropic thermal expansion methods deviate by 
0.03--0.17 kcal/mol. The main reason is the fact that the off-diagonals in the lattice tensor change as temperature 
increases, which is not captured in the 3D expansion approach. Methods using the Gr\"{u}neisen approach differ from 
those using the full Hessian calculations by between 0.05--0.09 kcal/mol.

    Ignoring thermal expansion is inappropriate for these crystal structures, but the difference between isotropic and
anisotropic methods is crystal dependent. Overall, thermal expansion can reduce the free energy computed in the 
harmonic approximation by 0.17--0.46 kcal/mol for methods that compute the full Hessian or 0.18--0.55 kcal/mol for 
Gr\"{u}neisen approaches. Isotropic expansion makes up the majority of this difference, up to 0.36 and 0.40 kcal/mol 
for full Hessian and Gr\"{u}neisen approaches, respectively. The anisotropic methods only reduce the free energy by 
0.01--0.04 kcal/mol for resorcinol and 0.02--0.23 kcal/mol for piracetam. The largest deviation between isotropic and 
anisotropic approaches comes from piracetam form III where 1D-GradAniso-QHA$\gamma$ is 0.23 kcal/mol lower in energy
than the isotropic Gr\"{u}neisen approaches. 

    QuasiAniso-QHA converges to high temperature free energies more similar to isotropic expansion rather than anisotropic
expansion. The quasi-expansion approach deviates between 0.02--0.23 kcal/mol from all fully anisotropic methods.
For piracetam form III and resorcinol form $\beta$ the 300 K free energy computed by QuasiAniso-QHA is actually
higher than GradIso-QHA. The lattice energy, $U(V)$, is always lower in energy for Quasi-Aniso-QHA than the 
isotropic methods (as it must be). However, we found that the vibrational contribution to the free energy made 
the quasi-anisotropic structure less favorable at higher temperatures, leading to minimization at different 
structural points. The quasi-anisotropic approach ranged between -0.006--0.03 kcal/mol of GradIso-QHA at 300 K, 
frequently being higher in free energy than GradIso-QHA at these temperatures, as compared to lower free energy, 
more accurate 1D-GradAniso-QHA and GradAniso-QHA methods (and their Gr\"{u}nisen variants). Further detail and 
analysis can be found in the Supporting Information (Section~\ref{section:Quasi-Aniso-failure}).

\subsection{Gibbs Free Energy Differences of Polymorphs}
    The Gibbs free energies of individual polymorphs are not physically relevant; the physically relevant observable 
that determines their stability is the Gibbs free energy difference between the polymorphs. It is often assumed that 
errors in the type of thermal expansion, finite size effects, anharmonic motions, and Gr\"{u}neisen parameter, and even 
to some extent the potential energy function used to model the crystal structure can be neglected because of 
cancellation effects in polymorph free energy differences. 

\begin{figure*}[!htb]
    \begin{center}
    \includegraphics[width=12cm]{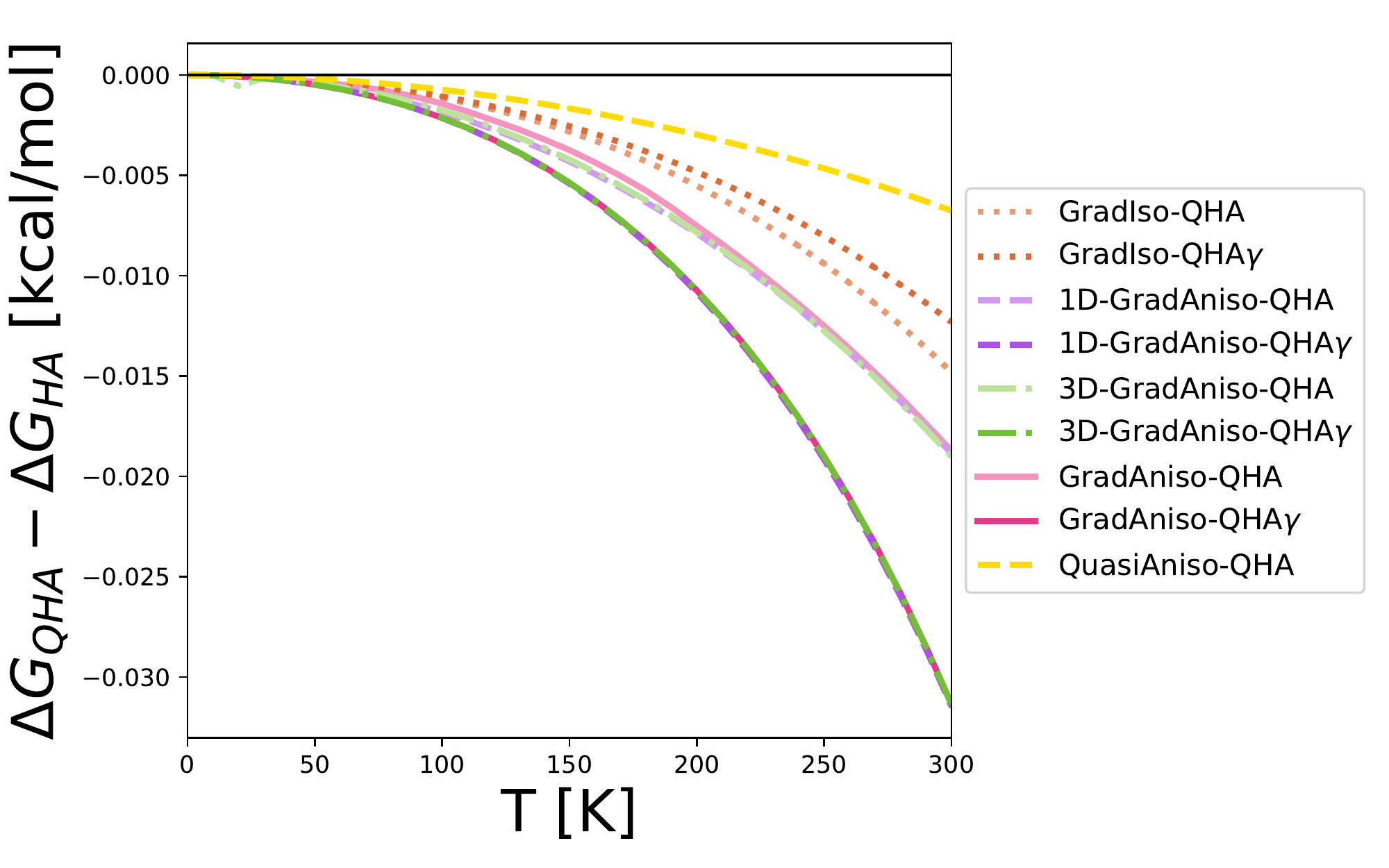}
    \includegraphics[width=12cm]{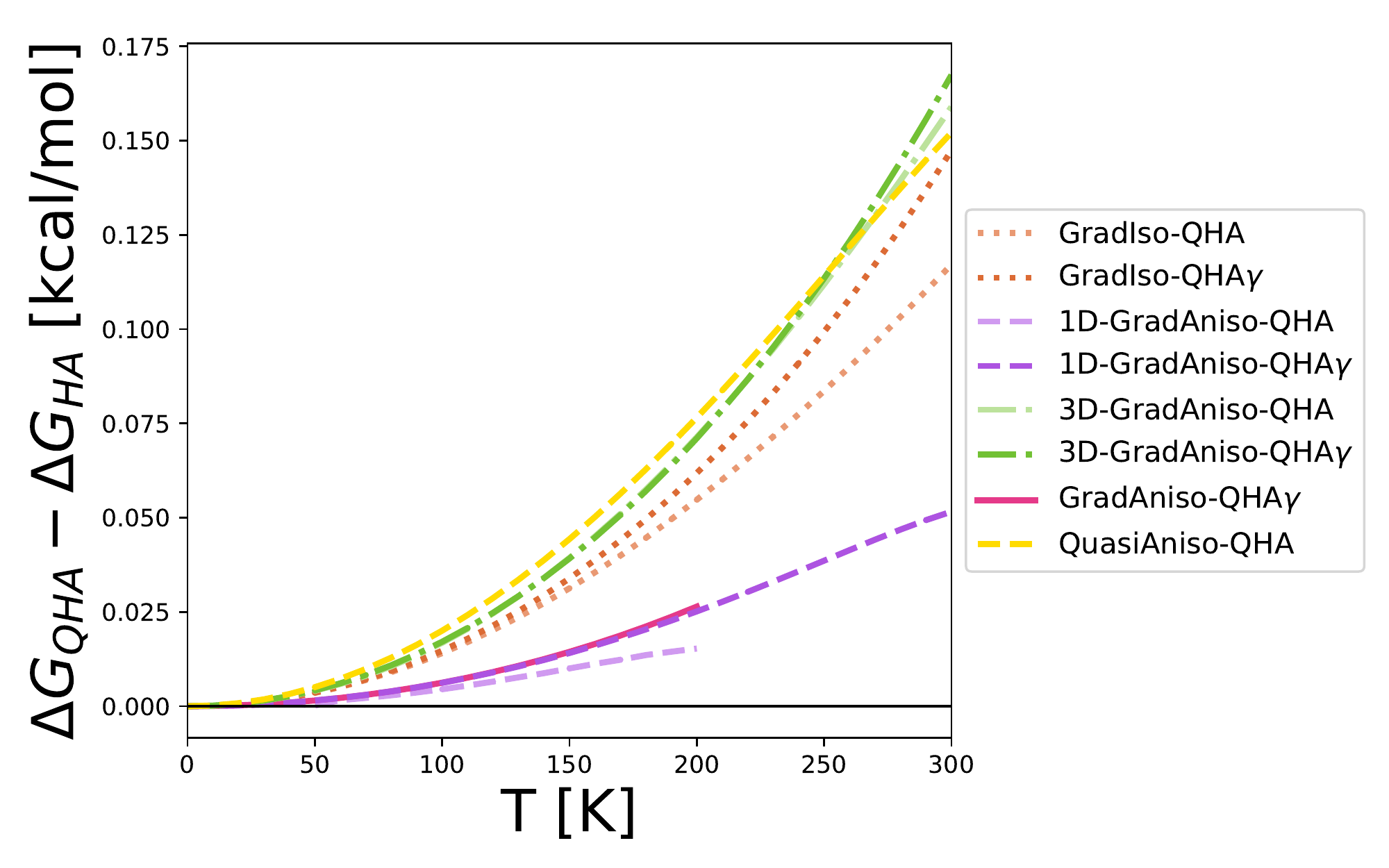}
    \end{center}
    \caption{Computed polymorph Gibbs free energies differences for the QHA methods shown as their deviation from the 
             harmonic approximation using eq.~\ref{equation:polymorph_diff} for resorcinol (top)
             and piracetam (bottom). The top graph shows that all QHA methods 
             predict the free energy differences within 0.02 kcal/mol of each other for resorcinol. The bottom graph 
             shows that all QHA methods predict the free energy differences with 0.12 kcal/mol for piracetam. Plots for 1D-GradAniso-QHA and GradAniso-QHA$\gamma$ are truncated and GradAniso-QHA omitted in piracetam because of lower temperature problems with restructuring on expansion (see Section~\protect{\ref{section:results_discontinuities}}).
    \label{figure:dDGvsT}}
  \end{figure*}

    We therefore desire to determine if errors in the type of thermal expansion and the use of the Gr\"{u}neisen parameter cancel 
for polymorph free energy differences. In fig.~\ref{figure:dDGvsT} the deviations of polymorph free energy differences for 
QHA are shown from HA. We quantify the deviations ($\Delta \Delta G_{i,j}$) by eq.~\ref{equation:polymorph_diff}, where 
$\Delta G^{i,j}$ is the free energy difference between polymorphs $i$ and $j$.
  \begin{eqnarray}
    \Delta \Delta G_{i,j} = \Delta G_{QHA}^{i,j} - \Delta G_{HA}^{i,j}
    \label{equation:polymorph_diff}
  \end{eqnarray}
Comparing the polymorph free energy differences, we see that all QHA methods compute $\Delta G_{i,j}$ within 0.02 kcal/mol 
or 0.12 kcal/mol from each other for resorcinol and piracetam respectively. Since there is such negligible 
difference between stepwise and gradient isotropic approaches, the results for StepIso-QHA and StepIso-QHA$\gamma$ are
not shown. The isotropic Gr\"{u}neisen approaches hav the largest deviations from other isotropic methods. The largest deviation 
due to use of the Gr\"{u}neisen approaches is 0.0025 kcal/mol for resorcinol and 0.025 kcal/mol for piracetam at 300 K. 

    Anisotropic expansion yields little change to the polymorph free energy differences for resorcinol, but provides a 
greater difference for piracetam, as shown in fig.~\ref{figure:dDGvsT}. For resorcinol, the polymorph free energy difference between isotropic approaches and 
unconstrained anisotropic expansion increases with temperature, but the overall deviation of 0.02 kcal/mol is 
relatively small. In contrast, the type of thermal expansion can significantly alter the final free energy difference 
for piracetam. Despite the re-structuring in piracetam form III we are able to show the results for 
GradAniso-QHA$\gamma$ up to 200K, prior to failure. At temperatures less than 200 K, we see that 1D-GradAniso-QHA$\gamma$
is nearly indistinguishable from GradAniso-QHA$\gamma$, implying that the 1D approach is a more accurate model
for anisotropic thermal expansion than the 3D approach. In particular, changes in the off-diagonal elements of the 
crystal tensor are responsible for a 0.08--0.12 kcal/mol change in $\Delta G_{i,j}$ from all other methods at 300 K 
for 1D-GradAnsio-QHA$\gamma$ and the other methods.

    The quasi-anisotropic QHA method computes the free energy differences of the polymorphs only to within 0.025--0.1 
kcal/mol of our most accurate method for the molecules tested, as shown in fig.~\ref{figure:dDGvsT}. For resorcinol, 
the quasi-anisotropic approach performs slightly worse than the isotropic methods relative to the GradAniso-QHA and 
GradAniso-QHA$\gamma$, but the energy gaps between all methods are less than 0.025 kcal/mol. For piracetam, the 
agreement between GradAniso-QHA$\gamma$ and 1D-GradAniso-QHA$\gamma$ at low temperatures strongly suggests that at 
300 K, 1D-GradAniso-QHA$\gamma$ is the most accurate method, and QuasiAniso-QHA deviates by 0.1 kcal/mol from this 
approach at that temperature.

\subsection{Lattice Expansion}
    Changes in the Gibbs free energy and free energy differences between different QHA approaches are due to the varying 
lattice geometries at high temperatures. We evaluate those changes by computing the percent change in the lattice 
vectors and angles (eq.~\ref{equation:lattice_expansion}) for all four crystals. We compare the percent lattice change 
at 300 K, $\% h(T=300\:K)$, where the percent expansion will be greatest.

\begin{figure*}[!htb]
    \begin{center}
    \includegraphics[width=8cm]{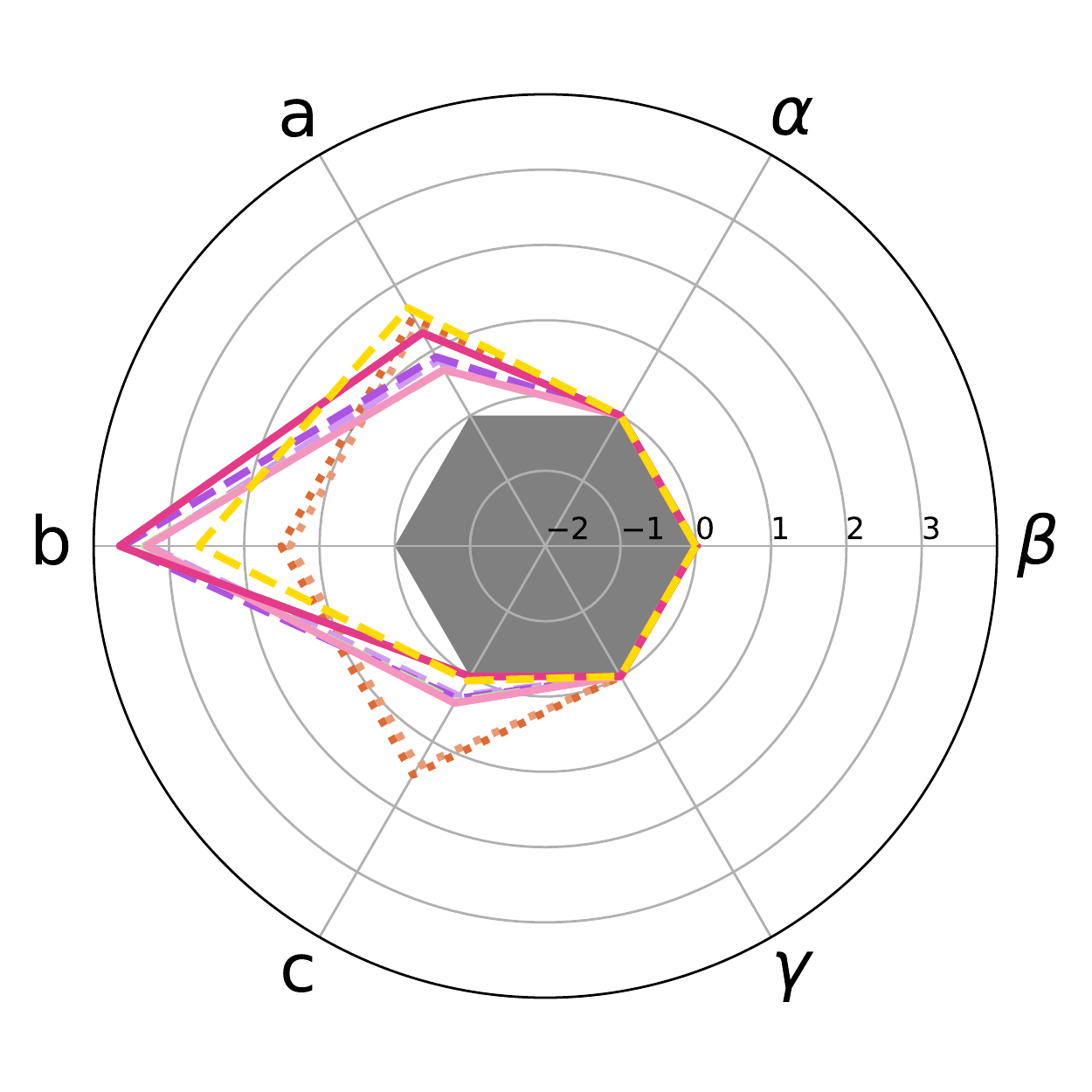}
    \includegraphics[width=8cm]{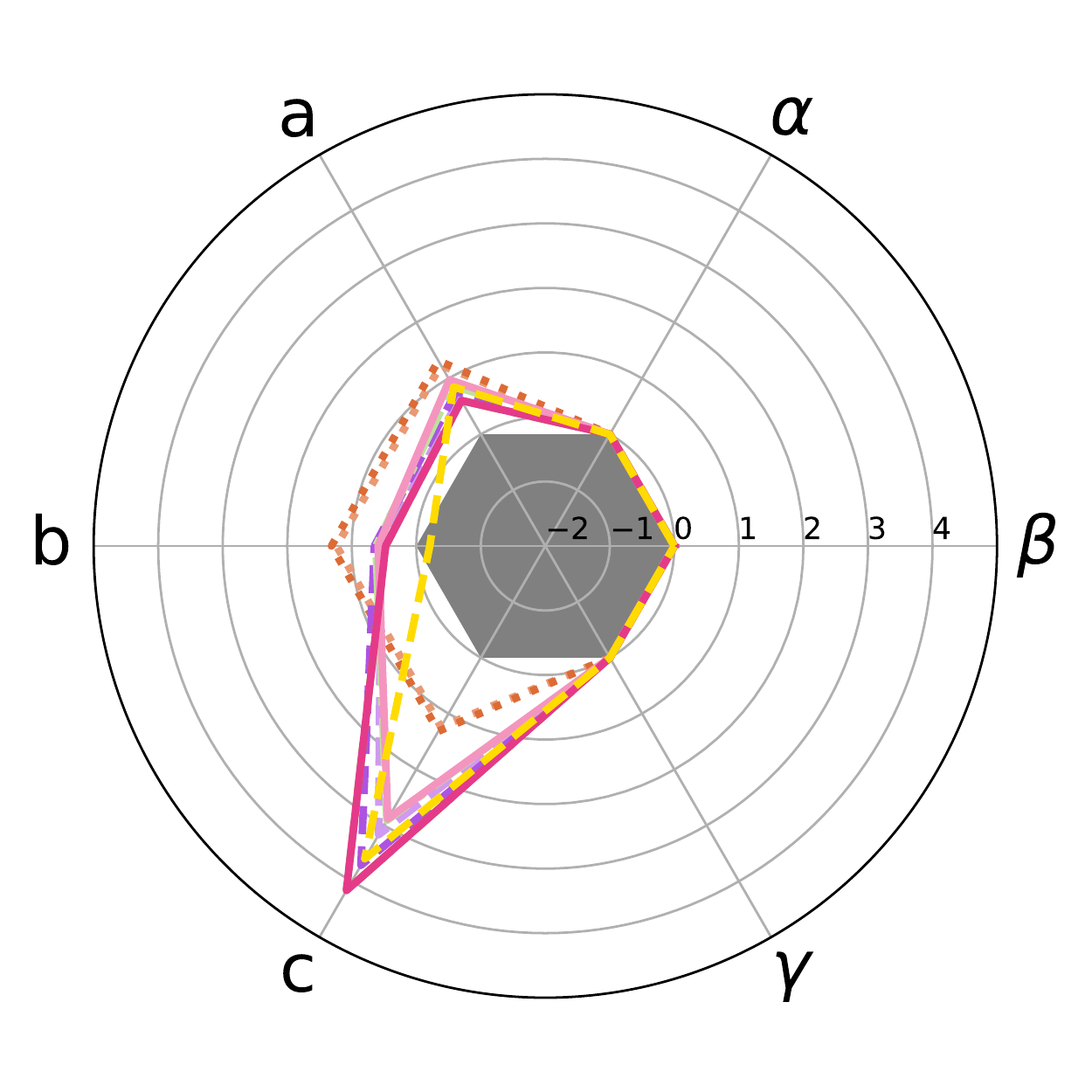} \\
    \includegraphics[width=8cm]{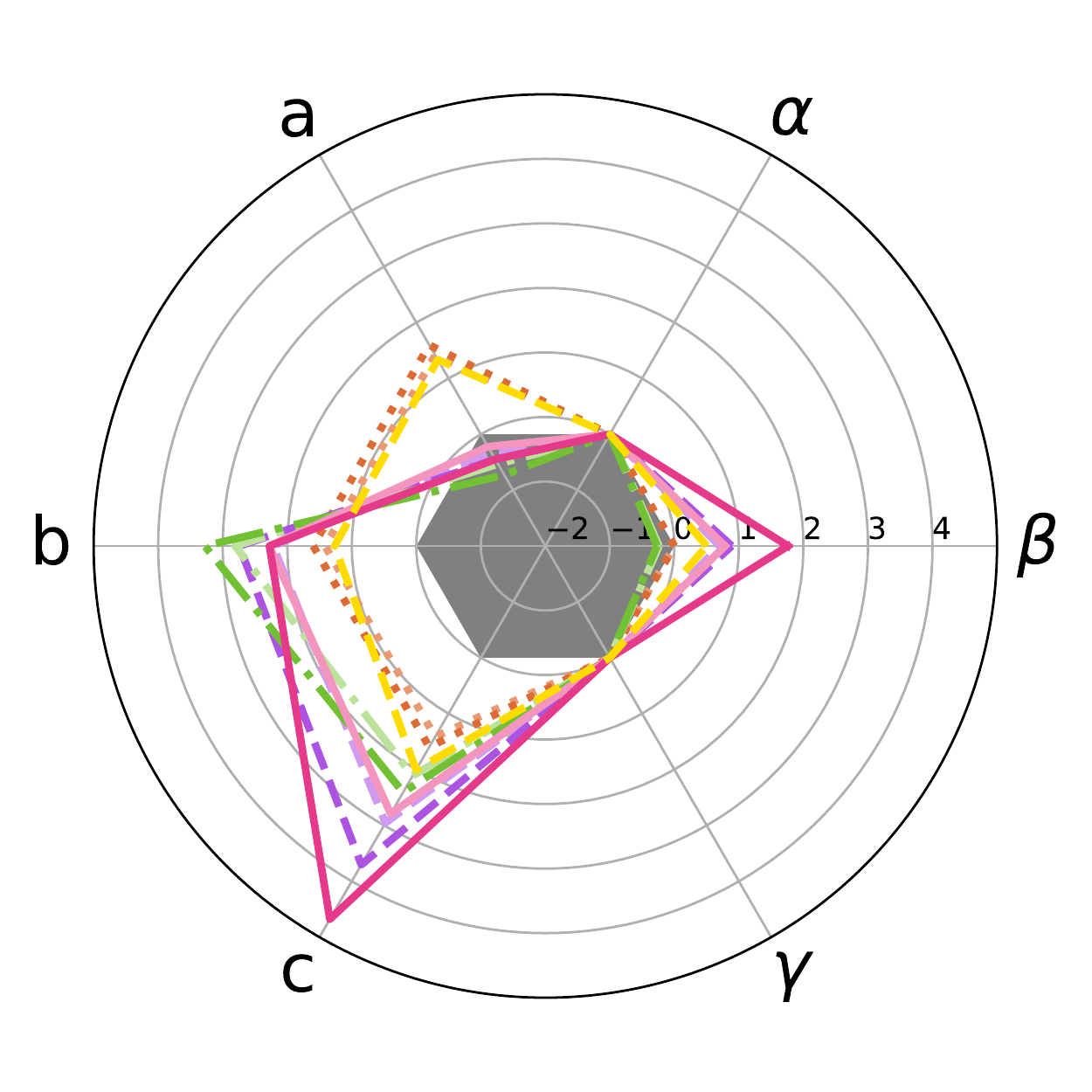}
    \includegraphics[width=8cm]{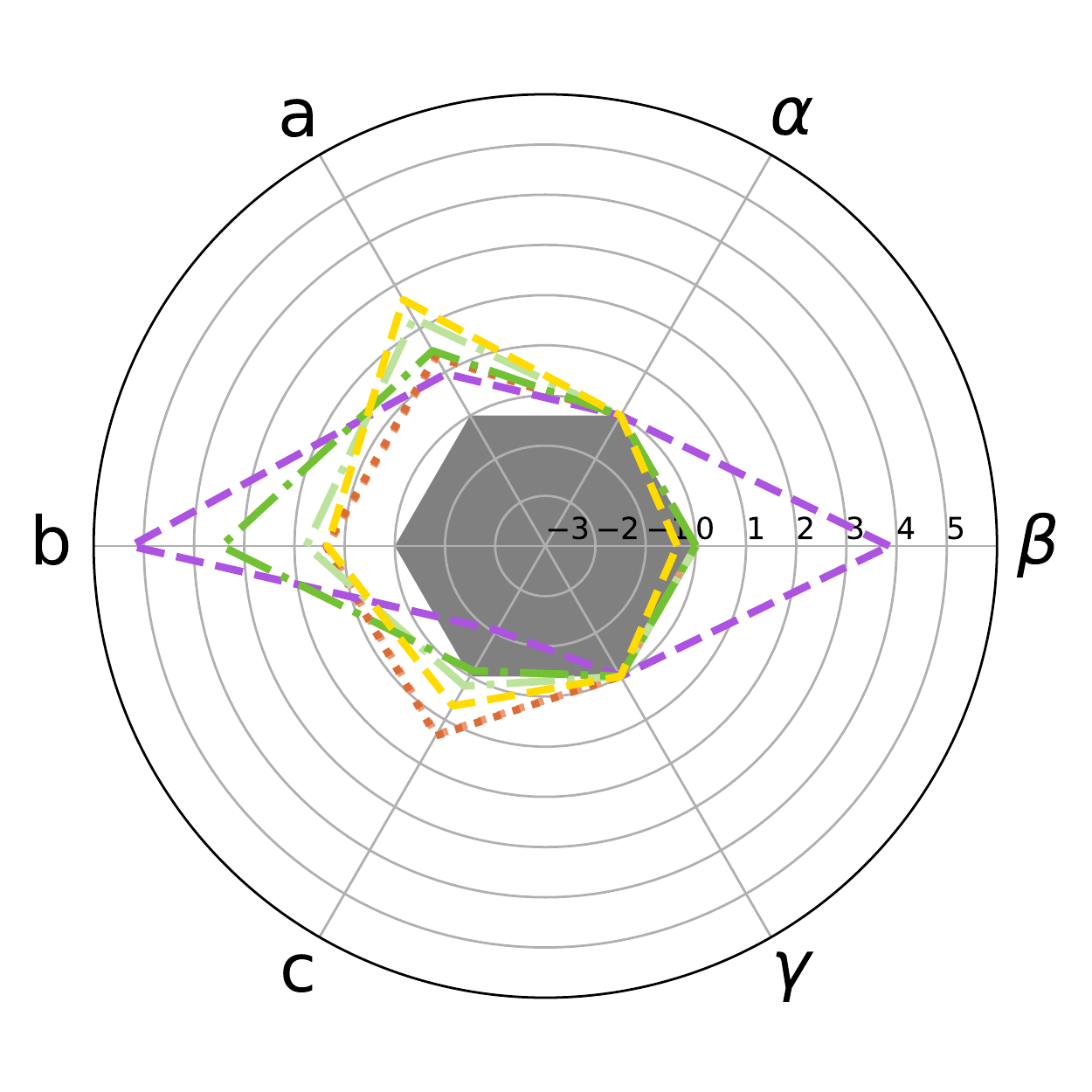} \\
    \includegraphics[width=16cm]{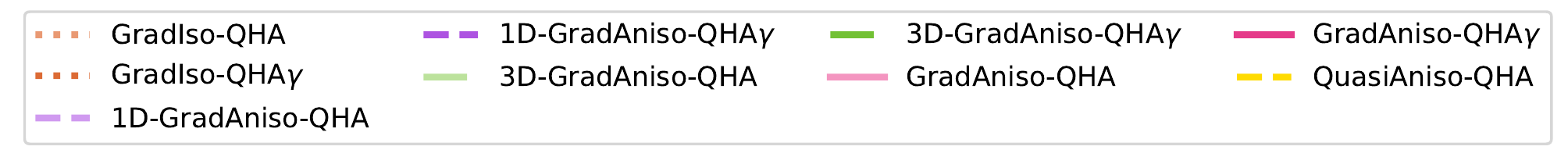}
    \end{center}
    \caption{Computed percent change of lattice parameters from 0 to 300 K for resorcinol form $\alpha$ (top left) and 
             $\beta$ (top right), as well as piracetam form I (bottom left) and III (bottom right). The white region
             reflects a percent increase with temperature, while the gray region is a percent decrease. The change in the 
             lattice parameters show distinct differences between isotropic and anisotropic expansion.
    \label{figure:Radar}}
  \end{figure*}

    The results for all eight gradient QHA methods in fig.~\ref{figure:Radar} show the distinct differences for how the 
crystals expand for isotropic and anisotropic thermal expansion. Stepwise approaches are not shown because they are 
within numerical error of the gradient approach. For each radar plot, the left shows the percent change in the
vectors and the right shows the percent change for the lattice angles. All percent changes are from 0 to 300 K as 
computed with eq.~\ref{equation:lattice_expansion}. If the points of the triangle lie within the white region there is 
a percent increase and the gray region signifies a percent decrease of the parameter. All isotropic methods have the 
same percent change in the lattice vectors and no change in the angles.
    All isotropic methods converge to the same lattice geometry at $T >$ 300 K within numerical error, as expected. The 
isotropic methods are plotted as the solid lines on the radar plots. Close examination of the four isotropic QHA 
methods shows that the predicted expansion of lattice parameters are visually indistinguishable across all four methods. 
The largest range of the percent expansion of lattice vectors is 0.19\% and that is for piracetam form I, which is 
small relative to the differences due to anisotropic expansion. Despite the errors in the Gibbs free energy when 
using the Gr\"{u}neisen approaches, all four methods converge to the same minimum energy lattice geometry.

    Isotropic expansion largely underestimates the changes in the lattice vectors due to thermal expansion, and also
misses important changes in the angles. For all four crystals, we can see that the triangles on the radar plots for the 
lattice vectors represent a completely different set of percent expansions from the isotropic methods. In all four 
crystals we see that one vectors expands about 2--4\% more than the isotropic methods. Alternatively, we can see that 
only one angle changes for both polymorphs of piracetam, but that percent expansion can be as large as a 4\% change 
in the angle, which supports the large energy differences seen between the methods.

    The quasi-anisotropic approach produces similar high temperature lattice geometries as the anisotropic approaches for
resorcinol, but not piracetam. For both crystal structures of resorcinol, the results with the quasi-anisotropic approach 
are somewhat 
different from the fully anisotropic methods, but do qualitatively capture the same expansion trends. This is contrasted 
with piracetam, where the quasi-anisotropic approach produces a high temperature lattice geometry that is much closer to 
isotropic expansion than to the anisotropic expansion. These results are only for a small data set, and thus may not be 
representative of results obtained with other molecules.

\subsection{Success of One-Dimensional Anisotropic Gradient QHA}
    Our one-dimensional anisotropic gradient approach produces results similar to the unconstrained anisotropic 
gradient method and is only slightly more computationally expensive than the gradient isotropic approach. For the free
energy, free energy differences, and lattice expansions we see that 1D-GradAniso-QHA and 1D-GradAniso-QHA$\gamma$ are 
indistinguishable from GradAnsio-QHA and GradAniso-QHA$\gamma$, respectively. Additionally, the discontinuity found 
in the unconstrained anisotropic expansion of piracetam form III is also found by 1D-GradAniso-QHA. While these results
are for only two pairs of polymorphs, the accuracy thus far is promising especially considering the speed of 
1D-GradAniso-QHA. In table~\ref{table:methods_NumSubRoutines} the number of geometry and optimizations for each method
are provided. We can see that our one-dimensional approach will be slightly slower than the isotropic method, but
requires much less computational time compared to the unconstrained approach.
  \begin{table*}[!htb]
  \begin{tabular}{ |c|c|c| }
    \hline
    \multicolumn{3}{ |c| } {Number of Sub-Routines}\\
    \hline
    Method & Number of Geometry Optimizations  &  Number of Hessians \\ \hline
    GradIso-QHA & 39 & 39 \\
    1D-GradAniso-QHA & 58--112  & 58--112 \\
    GradAnsio-QHA & 247--949  & 247--949  \\ \hline
    GradIso-QHA$\gamma$ & 39 & 3  \\
    1D-GradAniso-QHA$\gamma$ & 64--118 & 7 \\
    GradAnsio-QHA$\gamma$ & 253--955  & 7 \\
    \hline
  \end{tabular}
  \caption{For a select number of methods we show how many geometry optimizations and Hessian calculations are 
           required. We provide a range for the anisotropic methods because specific pieces of the lattice tensor
           may be zero, in which case the gradient at those points are not calculated. All of these results are for
           the we laid out previously in our methods section.
  \label{table:methods_NumSubRoutines}}
  \end{table*}

\section{Conclusions}
    We have presented a novel method for calculating both isotropic and anisotropic expansion of crystalline materials 
in the quasi-harmonic approximation. We compare our new gradient approach to established stepwise grid search methods 
for isotropic thermal expansion as well as quasi-anisotropic approaches and find that the stepwise and gradient approaches 
are essentially indistinguishable. Additionally, we demonstrate that our approach is easily extended to more complex 
anisotropic expansion and show how we can limit the degree of anisotropy sampled with the method.

    Our gradient method produces results within numerical error of the isotropic stepwise approach, but we do note that 
assuming the Gr\"{u}neisen parameter is constant can lead to non-negligible deviations in the free energy at high 
temperatures. The computed free energy at 300 K for stepwise and gradient isotropic approaches are within 0.0002 kcal/mol 
for all crystals. However, Gr\"{u}neisen approximation approaches can vary up to 0.04 kcal/mol at 300 K from methods 
that compute the full Hessian. Some of the error caused by the Gr\"{u}neisen approximation is canceled when looking at 
polymorph free energy differences. All four isotropic QHA methods compute the 300 K polymorph free energy differences
within 0.025 kcal/mol for both resorcinol and piracetam. 

    For these crystals, high temperature lattice geometries differ significantly with anisotropic expansion versus isotropic 
expansion, but not for quasi-anisotropic QHA. For isotropic expansion, the lattice vectors have equal percent expansions 
(1--2\%) from 0 to 300 K and the lattice angles remain fixed. For anisotropic expansion the lattice vectors change between 
-1--5\% and both crystals of piracetam exhibited a change in the $\beta$ angle. Quasi-anisotropic QHA matches our anisotropic
approaches to within 0.5\% expansion of each lattice parameter for resorcinol, but misses the anisotropic nature of the
crystals of piracetam and closely resembles the isotropic expansion trends.

    The effect of adding anisotropic expansion on the free energies is system dependent, but even for cases where it is 
important there is a cancellation of error in the free energy differences between polymorphs. Anisotropic expansion 
reduces the crystal free energy at 300 K by 0.01--0.23 kcal/mol relative to isotropic expansion because of the removal 
of the isotropic constraints. However, these changes in free energy resulted in changes in the free energy differences 
between polymorphs of only 0.01--0.12 kcal/mol compared to isotropic expansion, showing that a large amount of the error 
in the isotropic model is canceled in the relative energy differences between polymorphs. Anisotropic expansion is more 
important for piracetam than for resorcinol, with differences in free energy and free energy differences between 
polymorphs using anisotropic versus isotropic expansion for piracetam approximately twice those for resorcinol.

    Our thermal gradient approach demonstrates that minimization of the full free energy, and not just the lattice energy, 
is important when treating anisotropic expansion effects on high temperature lattice geometry and free energy. Including 
anharmonic effects fully is less relevant for polymorph free energy differences, however. Despite the fact that the free 
energies of the crystals at 300 K can differ by 0.02--0.23 kcal/mol quasi-anisotropic and anisotropic approaches, the free
energy differences are all within 0.025 kcal/mol. Overall, the quasi-anisotropic approach provides somewhat better insight into 
the crystal properties relative to isotropic expansion, but is problematic in at least some molecules. 
The introduction of our gradient approach allows us to perform a full free energy minimization of the crystal lattice with similar efficiency, at least in the case the 1D variant.

    A significant challenge when utilizing any lattice dynamic method is to assure that the method is sampling only 
one free energy minimum across the entire temperature range of interest. As we saw with piracetam, two anisotropic
QHA methods failed due to a re-structuring of the crystal upon expansion to higher than 200 K, a scenario which QHA is 
not designed to handle for more complex crystals where almost certainly multiple minima are accessible. Issues like 
this are likely to become more problematic with more complex molecules and increased number of variables in the 
optimization, as with 1D-GradAniso-QHA and GradAniso-QHA$\gamma$. Our best recommendation for identifying potential 
failures is to take more Runge-Kutta steps until $\frac{\partial G}{\partial y} = 0$ ($y = V$, $\lambda$, $C_{i}$) or 
$\frac{\partial l}{\partial T}$ quickly increases / oscillates, indicating some sort of structural discontinuity in the 
expansion.

    For the model systems presented in this paper, we have successfully shown that our gradient approach provides 
a reasonably fast way to perform QHA with an fully anisotropic thermal expansion model. In these test systems, 
we saw that the lattice parameters relaxed anisotropically because of the availability of lower free energy geometries. 
While the number of molecules studied here is small, it demonstrates the viability of the approach presented here and 
shows that further exploration of the effects of anisotropic expansion is necessary. 

\section*{Acknowledgments}
    The authors thank Eric Dybeck and Natalie Schieber for help with discussion regarding of the importance of 
anisotropic thermal expansion and practicality of implementing the gradient method. We also thank Graeme Day, Gregory 
Beran, and Aaron Holder for input of the manuscript prior to submission. Testing and development of our methods / code 
as well final results were performed on the Extreme Science and Engineering Discovery Environment (XSEDE), which is 
supported by National Science Foundation grant number ACI-1548562. Specifically, it used the Bridges system, which is 
supported by NSF award number ACI-1445606, at the Pittsburgh Supercomputing Center (PSC). Testing was also performed on 
Summit super computer, which is supported by the NSF grant OAC-1532235. This work was supported financially by NSF 
through the grant NSF-CBET 1351635.

\newpage
\newpage

\newpage
\newpage

\maketitle

    All results shown here and in the main text are for the symmetric unit cells. We present how we chose the various 
parameters required in the paper. These include:
  \begin{enumerate}
    \item Standard error of Tinker's vibrational frequencies
    \item Dependence of the Gr\"{u}neisen parameter on finite difference step size
    \item Stepwise isotropic QHA grid spacing
    \item Gradient finite difference step size for isotropic, anisotropic, and 1D-anisotropic expansion
    \item Comparison of Euler and Runge-Kutta step sizes
    \item Full polymorph free energies summarized in the paper
    \item Analysis of quasi-anisotropic lattice energies and free energies
    \item Modifications required for Tinker
    \item Details of calculation of the Gr\"{u}neisen parameter using the strain
    \item Matching vibrational modes
    \item Description of the constrained volume optimization procedure
  \end{enumerate}
We also show the polymorph free energy differences for each crystal up to 300 K, modifications we had to make to 
Tinker, and discussion of how we computed the strain tensor for the Gr\"{u}neisen parameter. All unit cells were 
retrieved from Cambridge Crystallographic Data Center. The corresponding reference codes are BISMEV03 for piracetam 
form I, BISMEV02 for piracetam from III, RESORA03 for resorcinol form $\alpha$ and RESORA08 for form $\beta$.

\section{Standard Error of Tinker's Vibrational Frequencies} \label{section:Tink_ViqFreq}
    Obtaining a highly accurate crystal vibrational spectra is both one of the most important and hardest parts of converging
on a numerically stable QHA result. We determine the sensitivity of the vibrational spectra from Tinker's {\tt vibrate} 
executable based on the tolerance used in the geometry optimization, {\tt minimize}, for the crystals studied. Aspects like the
Hessian tolerance and PME-grid spacing effect the vibrational sensitivity, the minimization criteria provides the greatest 
reduction in error. 

\begin{figure*} 
    \begin{center}
    \includegraphics[width=16cm]{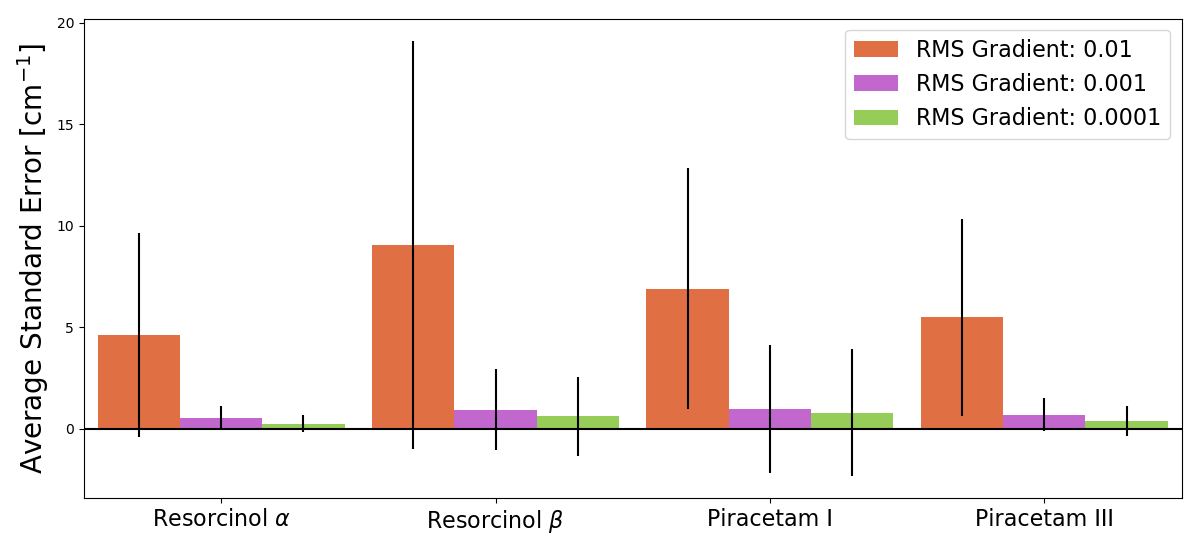}
    \end{center}
    \caption{A geometry optimization tolerance of $10^{-4}$ RMS gradient per atom reduces the sensitivity of the crystal
             vibrational spectra by an order of magnitude relative to a tolerance of $10^{-2}$. Error bars show that 
             individual modes may have larger standard error.
    \label{figure:Minimization_RMS}}
  \end{figure*}

    We compute the sensitivity by taking the 4 lattice minimum structures and isotropically expanding them by 
${0.001, 0.002, 0.003, 0.004}$\%$V$. Each structure was geometry optimized to an RMS gradient per atom tolerance of $10^{-N}$, 
where $N = 2, 3,$ and $4$  using {\tt minimize} and the vibrational spectra are calculated with {\tt vibrate}. We assume 
that within the 0.004\% $V$ change the individual vibrational modes should be approximately equal to one another another and
therefore can be fitted to a linear approximation with a slope of zero. This approximation is different than that made with
the Gr\"{u}neisen parameter since we are working with such small volume changes. We computed the standard error of a linear fit
to each vibrational mode versus volume, and report the average of all $3N$ vibrational modes ($N$ is the number of atoms per
unit cell). These results can be seen in fig.~\ref{figure:Minimization_RMS}.

    In the case of form $\beta$ resorcinol, the average standard error in the wavenumbers can be reduced
by an order of magnitude from $N = 2$ to $N = 4$. Using a geometry optimization of $10^{-4}$ RMS gradient per atom tolerance
sufficiently reduces the standard error and we see little reduction in error between $N = 3$ and $4$. We chose a geometry
optimization tolerance of $10^{-4}$.

    Converging the gradient to $10^{-4}$ in Tinker is numerically difficult and the subroutine occasionally exits 
without complete convergence. For minimizations that terminate prematurely,  we implemented a subroutine that shifts the 
atoms Cartesian coordinates in a random direction by less than $10^{-7}$ \AA~and attempts to re-minimize the 
structure. If the minimization does not converge within ten minimizations, the final structure, aborted in the $10^{th}$ 
minimization is used as the geometry optimized structure since the energies of all ten structures will be $<$ 0.001 
kcal/mol in difference. We found that about 5\% of the minimized structures could not reach a tolerance of $10^{-4}$ within 
the 10 rounds of our shake algorithm. If the desired tolerance is not reached a warning is output to the user, but the 
program continues using the structure output form the 10$^{th}$ minimization.

\section{Dependence of the \\ Gr\"{u}neisen Parameter on Finite Difference Step Size}
    The Gr\"{u}neisen parameters are solved numerically, so we want to determine how the Gr\"{u}neisen parameters as
we deviate from the lattice minimum structure. For all Gr\"{u}neisen step sizes, we only ran test for piracetam, but
the same response is expected for resorcinol.

\subsection{Isotropic Gr\"{u}neisen Parameter}
    As stated in the main paper, the isotropic Gr\"{u}neisen parameters are solved using a forward finite difference approach.
The numerical solution is shown in the following equation:
  \begin{eqnarray}
    \gamma_{k} = -\frac{\ln({\omega_{k}(V + \Delta V)}) - \ln({\omega_{k}(V)})}{\ln({1 + \frac{\Delta V}{V}})} 
    \label{equation:Gru_iso_solution} 
  \end{eqnarray}
To solve eq.~\ref{equation:Gru_iso_solution} we need to know the vibrational spectra of our lattice minimum structure
and one isotropically expanded structure. To determine the appropriate value of $\Delta V / V$ for that expanded 
structure, we will look at how the Gr\"{u}neisen parameters deviate from their value when $\Delta V / V \rightarrow 0$, where 
QHA is accurate. We plot the values of eq.~\ref{equation:Iso_Gru_deviation}, which is a unitless quantity, for varying 
$\Delta V / V$ to determine the appropriate step size. From this, we can determine how much the average Gr\"{u}neisen
parameter deviates from the low $\Delta V$ limit and look at a confidence interval for the deviations for all individual modes.
  \begin{eqnarray}
    \sum_{k=1}^{N} \frac{\gamma_{k} - \gamma_{k,0}}{N}
    \label{equation:Iso_Gru_deviation}
  \end{eqnarray}
$N$ is the total number of vibrational modes, $\gamma_{k}$ is the $k^{th}$ modes Gr\"{u}neisen parameter solved using
$\Delta V / V$, and $\gamma_{k,0}$ is the same modes Gr\"{u}neisen parameter as $\Delta V / V \rightarrow 0$. The
value of $\Delta V / V \rightarrow 0$ is actually at $\Delta V / V = 10^{-3}$, any lower value is numerically unstable 
as determined by the standard error in the angular frequencies.
  \begin{figure*}
    \begin{center}
    \includegraphics[width=8cm]{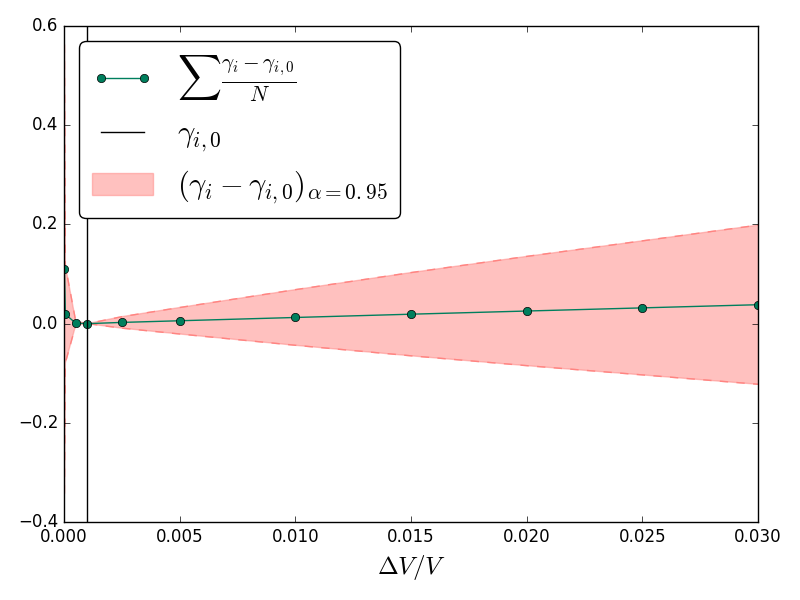}
    \includegraphics[width=8cm]{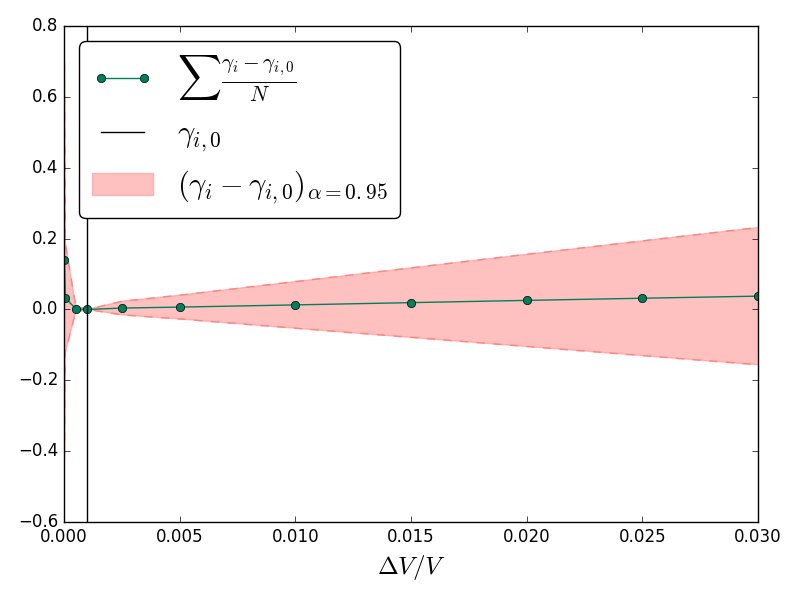}
    \end{center}
    \caption{Deviations in the isotropic Gr\"{u}neisen parameters for piracetam polymorph I (left) and III (right).
             The line is the average change in all of the parameters and the shaded region is the 95\% confidence
             interval of all individual modes. Small step sizes produces unstable results and large step sizes does 
			 not affect the average deviations, but can greatly impact individual modes.
    \label{figure:bismev_IsoGRU}}
  \end{figure*}

    The results in fig. \ref{figure:bismev_IsoGRU} show the change in the Gr\"{u}neisen parameter as we deviate from 
$\Delta V / V = 10^{-3}$ is small as shown by the green line, but that is not the case for individual modes. The 95\%
confidence interval shown in red shows that the individual Gr\"{u}neisen parameters can deviate 
up to 0.5 (unitless) from the Gr\"{u}neisen parameters for $\Delta V / V = 0$, which can be significant for the majority of 
Gr\"{u}neisen parameters that are localized around zero. For small step sizes, $\Delta V / V < 10^{-3}$, the solution for the 
Gr\"{u}neisen parameter is numerically unstable because of the standard error in Tinker's vibrational spectra. A step size of 
$\Delta V / V = 1.5\times10^{-3}$ is a small enough step size to reduce deviations of the Gr\"{u}neisen parameters from our 
limit and also assures us that numerical instabilities due to small step sizes are avoided.

\subsection{Anisotropic Gr\"{u}neisen Parameter}
    The anisotropic Gr\"{u}neisen parameter is also solved using a forward finite difference approach for the six different 
strains (3 normal and 3 shear strains), seen in eq.~\ref{equation:Gru_aniso_solution}.
  \begin{eqnarray}
    \gamma_{k,i} = - \frac{\ln({\omega_{k}(\eta_{i})}) - \ln({\omega_{k,0}})}{\eta_{i}}
	\label{equation:Gru_aniso_solution}
  \end{eqnarray}
We will use a similar formalism as the isotropic case to look at deviation in the Gr\"{u}neisen parameters:
  \begin{eqnarray}
    \sum_{k=1}^{N} \frac{\gamma_{k, \eta_{i}} - \gamma_{k, \eta_{i}, 0}}{N}
    \label{equation:Aniso_Gru_deviation}
  \end{eqnarray}
Where $\gamma_{k, \eta_{i}}$ is the Gr\"{u}neisen parameter for mode $k$ due to strain $\eta_{i}$ and 
$\gamma_{k, \eta_{i}, 0}$ is the parameter solved for $\eta \rightarrow 0$.
  \begin{figure*}
    \begin{center}
    \includegraphics[width=12cm]{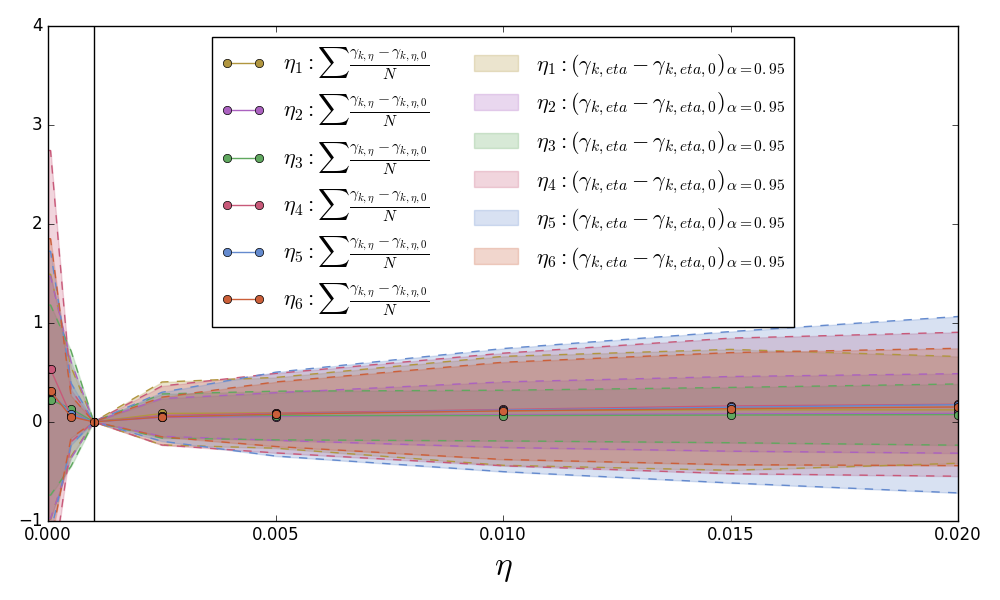}\\
    \includegraphics[width=12cm]{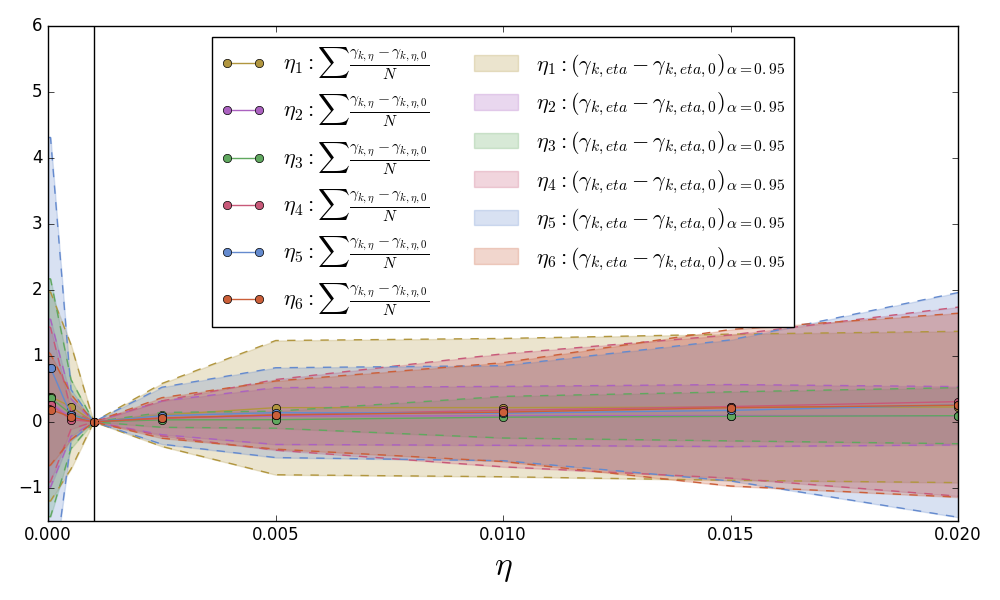}
    \end{center}
    \caption{Deviations in the anisotropic Gr\"{u}neisen parameters for piracetam polymorph I (top) and II (bottom).
             The lines are the average change in all of the parameters and the shaded region is the 95\% confidence
             interval of all individual modes. The anisotropic Gr\"{u}neisen parameters deviate by an order of magnitude 
             due to the step size in comparison to the isotropic case. Using a small step size is very important to 
             model the free energy accurately at low temperatures.
    \label{figure:bismev_AnisoGRU}}
  \end{figure*}

    Figure \ref{figure:bismev_AnisoGRU} shows the results for both crystals and deviations of the 6 strain Gr\"{u}neisen
parameters. Once again, the average deviation is close to zero, but the highlighted 95\% confidence interval is an order 
of magnitude larger than the deviations in the isotropic Gr\"{u}neisen parameters. While anisotropic parameters are 
larger than the isotropic parameters, they are not an order of magnitude larger. This highlights the need to use a small 
step size for $\eta_{i}$ in eq.~\ref{equation:Gru_aniso_solution}, which is why we will use a step size of 
$\eta_{i} = 2.5 \times 10^{-3}$.

\section{Stepwise Isotropic QHA Grid Spacing} \label{section:StepIso_Spacing}
    To assure that the stepwise methods are converging onto correct structure minimizing the free energy, we want to 
assure that there is sufficiently fine sampling in the volume between the maximum expanded and compressed structures. 
To do this, we have run StepIso-QHA$\gamma$ with varying values of $\Delta V / V$ and compare the convergence of 
$G(T=300K)$ and smoothness of $V(T)$. The effects of sampling points will effect both StepIso-QHA and StepIso-QHA$\gamma$ 
in the same manner, so we use StepIso-QHA$\gamma$ to reduce the computational expense.

\begin{figure*}
    \begin{center}
    \includegraphics[width=8cm]{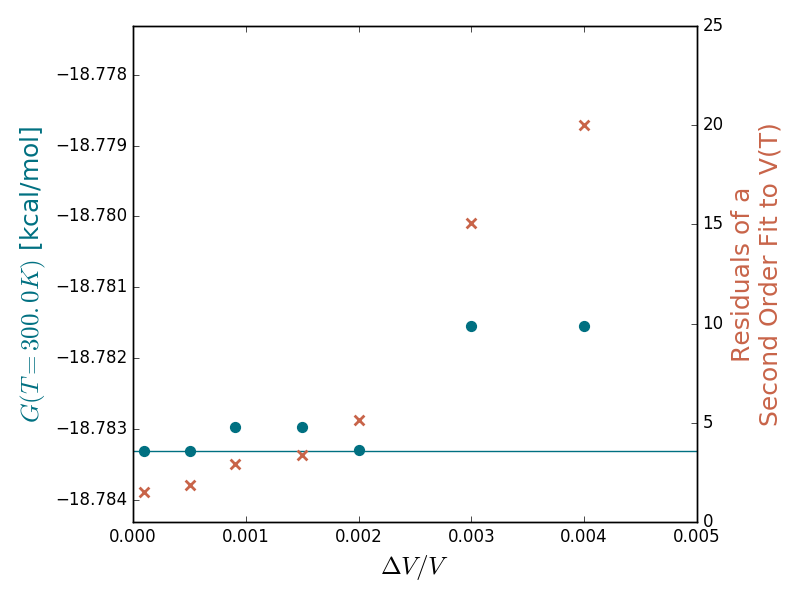}
    \includegraphics[width=8cm]{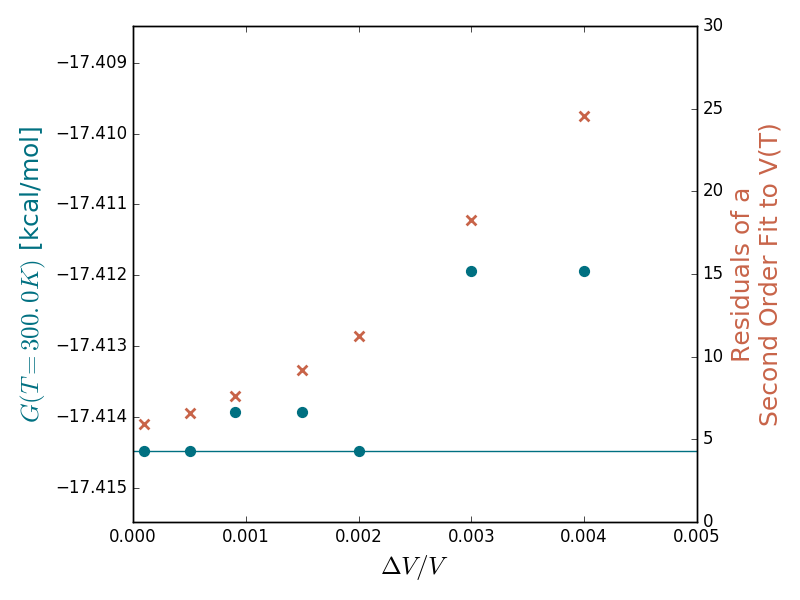}
    \end{center}
    \caption{Step sizes between adjacent isotropically expanded structures for stepwise isotropic QHA to show the
             convergence of $G(T)$ and smoothness of $V(T)$ for piracetam polymorph I (left) and III (right).
             The step sizes chosen here, have little influence on the free energy convergence at 300K, but have a large
             influence on the smoothness of $V(T)$. The teal line is the minimum Gibbs free energy computed for all step sizes.
    \label{figure:bismev_SiQ_VStepFrac}}
  \end{figure*}

    In Figure~\ref{figure:bismev_SiQ_VStepFrac} the results for both crystals of piracetam are reported. The left y-axis 
reports the $G(T=300K)$, the right y-axis reports the residuals of a second order fit to $V(T)$, and the x-axis has 
varying values of $\Delta V / V$ for StepIso-QHA$\gamma$. For all step sizes chosen we see that there are little 
deviations in $G(T=300K)$, the largest deviation is for piracetam form II with a deviation from the minimum energy (teal 
line) of 0.002 kcal/mol. The largest influence on the step size is the smoothness of the fit to $V(T)$, which becomes 
larger with the value of $\Delta V / V$. Based off of these figures, we have decided that $\Delta V / V = 10^{-3}$ is 
sufficient to converge onto the true isotropic QHA minimum for both stepwise methods. 

\section{Gradient Finite \\ Difference Step Size} \label{section:Gradi_step_size}
    As stated in the main paper, the best method for determining the finite step size for computing any of the 
gradients is to pick a step size that increased the potential energy of the lattice minimum structure by 0.0005 
kcal/mol. This cut-off will change depending on the numerical accuracy of the program used to compute the wavenumbers
and potential energy, and to perform minimizations. In our program, the user can input some value $f$ or the program
can determine the best value of $f$, where $f$ is defined as:
  \begin{eqnarray}
    f = \frac{\Delta y}{y_{0}}
  \end{eqnarray}
Where $y$ can be $V$ or $C_{i}$; $y_{0}$ is the value of $y$ at the lattice minimum structure; and
$\Delta y$ is the finite step size used to solve the thermal gradient ($\frac{\partial y}{\partial T}$). Further context
of $\Delta y$ is in the following sections.

    For each method (isotropic, anisotropic, and 1D-anisotropic) we perform an initial search for a value of $\Delta y$ where
$\Delta U > $ 0.0005 kcal/mol. Here, $\Delta U$ is:
  \begin{eqnarray}
    \Delta U = U(y_{0} + \Delta y) - U(y_{0})
  \end{eqnarray}
We do this for values of $f = 10^{-N}$ and $5\times10^{-N}$ for values of $N = 5, 4, 3, 2,$ and $1$. We use the first
value of $\Delta y$ that exceeds our cut-off for Tinker and that value for $\Delta y$ is used to compute every gradient in the run.

\subsection{Isotropic QHA}
\begin{figure}
    \begin{center}
    \includegraphics[width=8cm]{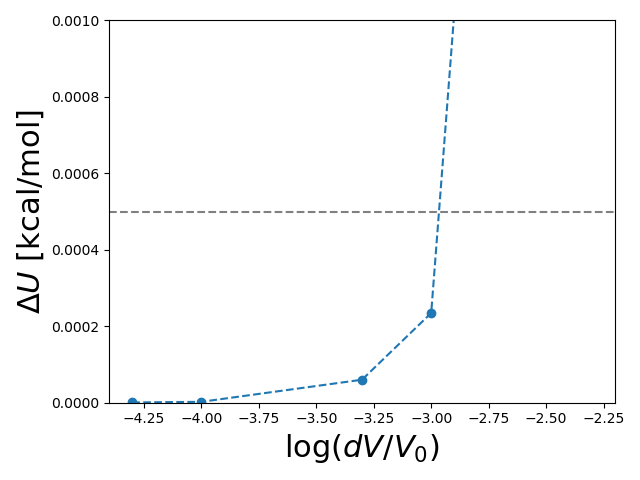}
    \end{center}
    \caption{Changes in the potential energy of the lattice minimum structe of piracetam form I, when changing the isotropic
             volume by varying values of $f = \Delta V / V_{0}$. The energy exceeds the required cut-off when $f = 5\times10^{-3}$.
    \label{figure:bismev_p1_U_dV}}
  \end{figure}

    As stated in the main paper, we have used a finite difference approach to solve eq.~\ref{equation:Gradient_iso}.
  \begin{eqnarray}
    \frac{\partial V}{\partial T} = \frac{(\frac{\partial S}{\partial V})}{(\frac{\partial^{2} G}{\partial V^{2}})} 
    \label{equation:Gradient_iso}
  \end{eqnarray}
Where the numerator and denominator are solved with a central finite difference approach:
  \begin{eqnarray}
    \frac{\partial S}{\partial V} = \frac{S(V + \Delta V) - S(V - \Delta V)}{2 \Delta V}
  \end{eqnarray}
  \begin{eqnarray}
    \frac{\partial^{2} G}{\partial V^{2}} = \frac{G(V + \Delta V) - 2G(V) + G(V - \Delta V)}{( \Delta V)^{2}}
  \end{eqnarray}
Fig.~\ref{figure:bismev_p1_U_dV} shows the results for piracetam form I for how $\Delta U$ changes as a function of $f$.
We can see that the first step to exceed our cut-off is $f = 5\times10^{-3}$. Therefore, we will $\Delta V$ for all gradients
will be set to $V_{0} \times 5\times10^{-3}$. For all four crystals the value of $f$ was set to $5\times10^{-3}$.

\subsection{Anisotropic QHA}
    We also solved for the anisotropic gradients, eq.~\ref{equation:Gradient_aniso}, with a finite difference approach. 
  \begin{eqnarray}
    \frac{\partial \boldsymbol{C}}{\partial T} = \left(\frac{\partial^{2} G}{\partial \boldsymbol{C}^{2}}\right)^{-1} \frac{\partial S}{\partial \boldsymbol{C}}
    \label{equation:Gradient_aniso}
  \end{eqnarray}
The second derivative of the free energy with the strain is a $6 \times 6$ matrix, where the diagonals are solved for $C_{i}$ as:
  \begin{eqnarray}
    \frac{\partial^{2} G}{\partial C_{i}^{2}} = \frac{1}{\Delta C_{i}^{2}} \left(G(T, \boldsymbol{C} + \Delta C_{i}) - 2G(T,\boldsymbol{C}) \right. \\ 
                \left. + G(T,\boldsymbol{C} - \Delta C_{i}\right)
  \end{eqnarray}
The off-diagonals of this matrix are symmetric and solved for $C_{i}$ and $C_{j}$ where $i \ne j$ as:
  \begin{equation*}
    \begin{aligned}
    \frac{\partial^{2} G}{\partial C_{i} \partial C_{j}} = \frac{1}{\Delta C_{i} \Delta C_{j}} (G(T, \boldsymbol{C} + \Delta C_{i} + \Delta C_{j}) \\
      - G(T, \boldsymbol{C} + \Delta C_{i} - \Delta C_{j}) \\
      - G(T,\boldsymbol{C} -  \Delta C_{i} + \Delta C_{j}) \\
      + G(T,\boldsymbol{C} -  \Delta C_{i} - \Delta C_{j}))
    \end{aligned}
  \end{equation*}
Lastly, the partial derivative of the entropy due to $\Delta C_{i}$ is solved as:
  \begin{eqnarray}
    \frac{\partial S}{\partial C_{i}} = \frac{S(T, \boldsymbol{C} + \Delta C_{i}) - S(T, \boldsymbol{C} - \Delta C_{i})}{2 \Delta C_{i}}
  \end{eqnarray}

   Similarly to the isotropic expansion we can find the values for $\Delta C_{i}$ ($i = 1, 2, ..., 6$) by determining the 
value of $f$ that exceeds our energy cut-off. Values for $f$ are shown in the following table:
  \begin{table*}[H]
  \begin{tabular}{ |l|c| }
    \hline
    \multicolumn{2}{ |c| } {GradAniso-QHA Finite Step Sizes}\\
    \hline
    Crystal &  $f$ \\ \hline
    Resorcinol $\alpha$ & \{$5\times10^{-3}, 5\times10^{-3}, 5\times10^{-3}, 0, 0, 0$\} \\
    Resorcinol $\beta$ &  \{$5\times10^{-3}, 5\times10^{-3}, 5\times10^{-3}, 0, 0, 0$\} \\
    Piracetam I &  \{$5\times10^{-3}, 5\times10^{-3}, 5\times10^{-3}, 0, 5\times10^{-1}, 0$\} \\
    Piracetam III &  \{$5\times10^{-3}, 5\times10^{-3}, 5\times10^{-3}, 0, 5\times10^{-2}, 0$\} \\
    \hline
  \end{tabular}
  \end{table*}

\subsection{1D-Anisotropic QHA}
    One dimensional is treated similarly to isotropic expansion, except $V$ is replaced with $\lambda$, but all
gradients are solved with a central finite difference approach. Based off of our definition $\lambda$ at the lattice
minimum structure is 0, so we have to re-define $f$ as $f = \Delta \lambda$. We can then determine the most appropriate value
of $f$ by determine when our energy cut-off is met. For all four crystals we found that the the value of $f = 5\times10^{-3}$.

\section{Euler and Runge-Kutta Step Sizes} \label{section:Euler_RK}
\begin{figure*}
    \begin{center}
    \includegraphics[width=8cm]{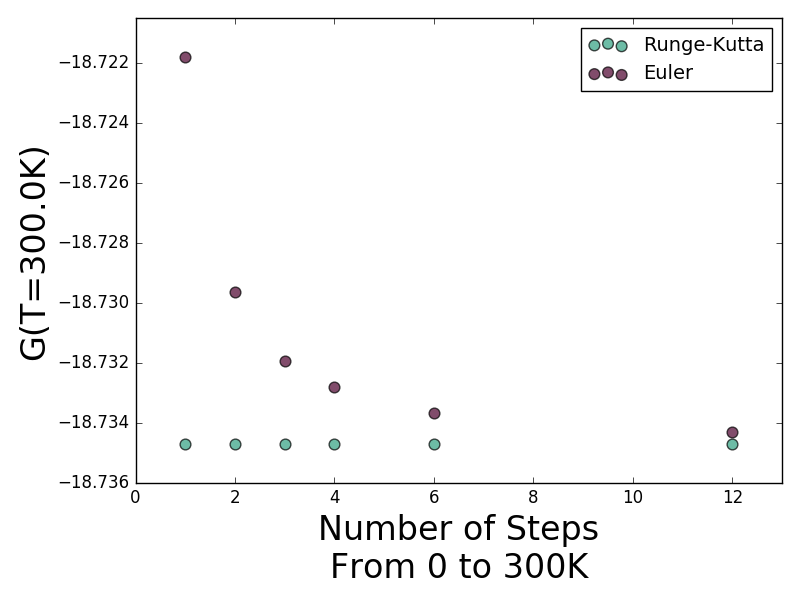}
    \includegraphics[width=8cm]{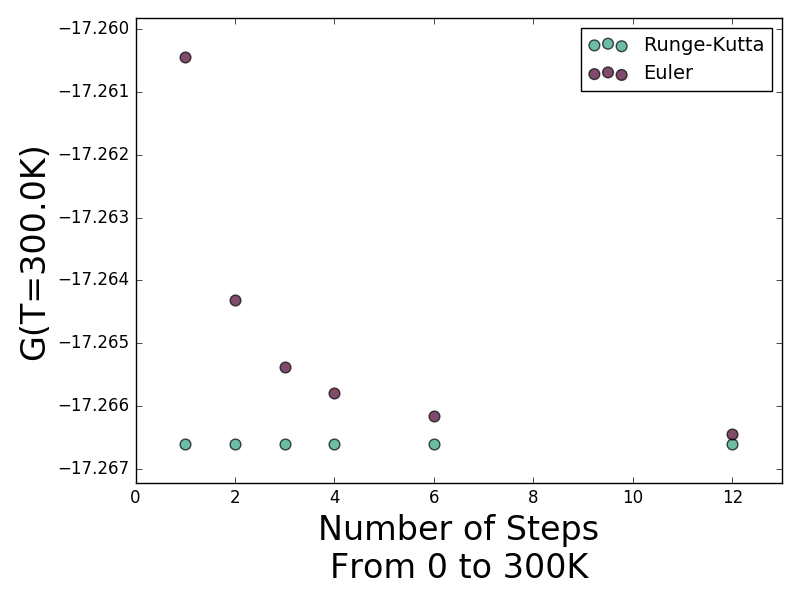}
    \end{center}
    \caption{Comparison of converged Gibbs free energy at 300K for Euler and $4^{th}$ order Runge-Kutta at varying step
             sizes for GradIso-QHA$\gamma$ on piracetam form I (left) and III (right). There is little difference 
             between 1 and 12 steps of Runge-Kutta, while Euler is greatly improved, but still does not converge on the 
             Runge-Kutta Result.
    \label{figure:bismev_iRKEU}}
  \end{figure*}

    A $4^{th}$ order Runge-Kutta is sufficiently cheap and numerically stable to use for GradIso-QHA$\gamma$ compared
to a Euler method. We have carried out Euler and $4^{th}$ order Runge-Kutta numerical methods for GradIso-QHA$\gamma$ 
with varying step sizes. We have chosen to use the Gr\"{u}neisen parameter because the results should be similar to 
GradIso-QHA and saves computational cost. In figure \ref{figure:bismev_iRKEU} the Gibbs free energy at 300K for both 
Euler and Runge-Kutta are compared for the varying number of steps between 0 and 300K to test the convergence of the 
results.

    In all cases, $G(T=300K)$ for Runge-Kutta is numerically indistinguishable for the varying number of steps used while the
Euler approach converges on to the same result as the number of steps increases. With 12 Euler steps, the results still 
have not fully converged onto the Runge-Kutta results and the 12 step Euler approach is 3$\times$ more expensive than a
single Runge-Kutta step. We have decided that a $4^{th}$ order Runge-Kutta approach is the better choice and, with 3 
steps from 0 to 300K (step size of 100K) having good performance. While similar results can be achieved with one step size 
for Runge-Kutta, we want to be certain that we have converged on the isotropic volume that minimizes the free energy at 
intermediate  points between 0 and 300K as well.

    Due to the computational expense of the anisotropic gradient methods, we were unable to conduct a full Euler and Runge-
Kutta analysis for anisotropic expansion. Since the results of isotropic expansion using the Runge-Kutta approach are stable, 
we have decided to perform anisotropic expansion with a three Runge-Kutta steps. If more steps were required, we stated so in
the main paper.

\section{Full Polymorph Free Energies} \label{section:polymorph_free_energies}
\begin{figure*}
    \begin{center}
    \includegraphics[width=8cm]{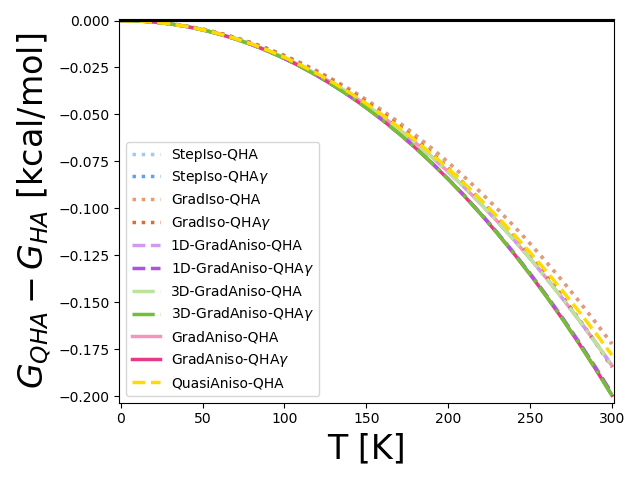}
    \includegraphics[width=8cm]{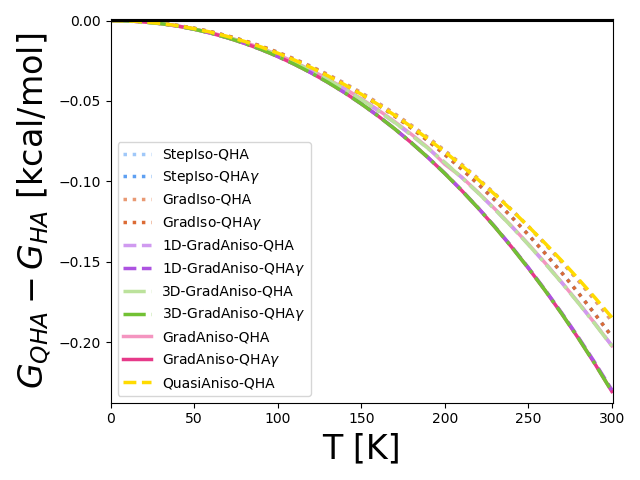} \\
    \includegraphics[width=8cm]{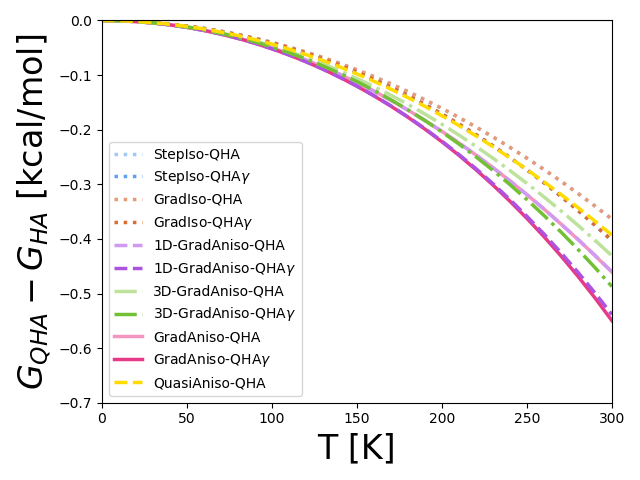}
    \includegraphics[width=8cm]{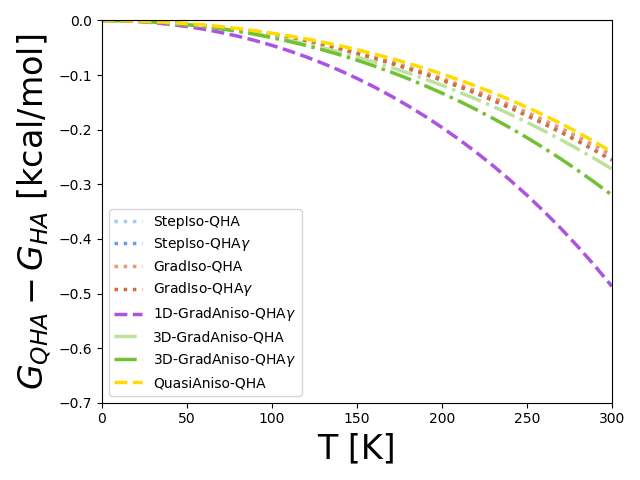}
    \end{center}
    \caption{Deviations of the free energy of QHA methods from HA for resorcinol form $\alpha$ (top left), $\beta$ (top
             right, and piracetam form I (bottom left) and III (bottom right). All methods show that the largest 
             deviations occur at high temperatures and the relative rankings in methods are 
             $G_{HA} > G_{Iso.-QHA} \geq G_{Aniso.-QHA}$.
    \label{figure:dGvsT}}
  \end{figure*}

    The plots in fig.~\ref{figure:dGvsT} are the complete results for polymorph free energies summarized in the main paper. As stated in the main 
paper, the largest deviations in all methods are at 300 K. At all temperatures, the isotropic methods produce similar 
results and the free energy ranking $G_{HA} > G_{Iso.-QHA} \geq G_{Aniso.-QHA}$ holds true.
  
 \section{Analysis of Quasi-\\anisotropic Lattice Energies and Free Energies} \label{section:Quasi-Aniso-failure}
\begin{figure*}[!htb]
    \begin{center}
    \includegraphics[width=8cm]{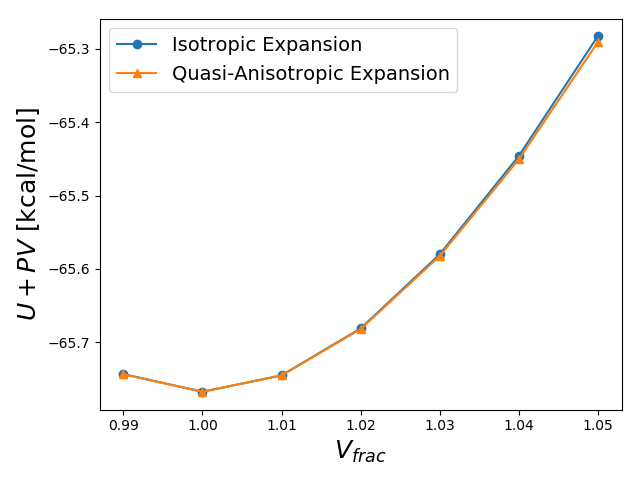}
    \includegraphics[width=8cm]{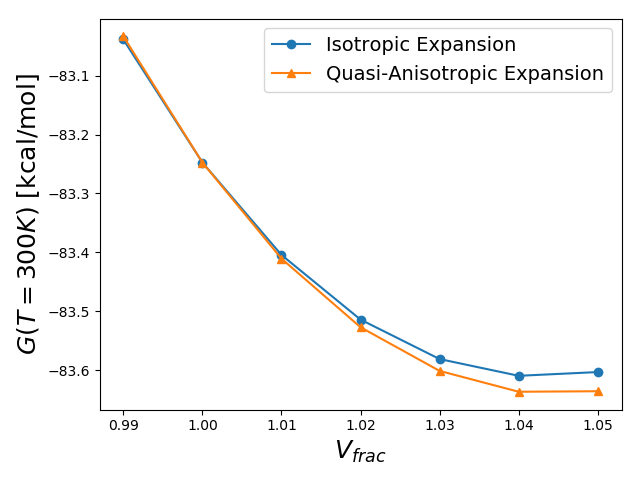}
    \end{center}
    \caption{Lattice enthalpy (right) and Gibbs free energy at 300 K for piracetam form I for varying volume fractions
             using isotropic expansion and quasi-anisotropic expansion.
    \label{figure:Quasi-Aniso_GvsV_bismev1}}
  \end{figure*}

\begin{figure*}
    \begin{center}
    \includegraphics[width=8cm]{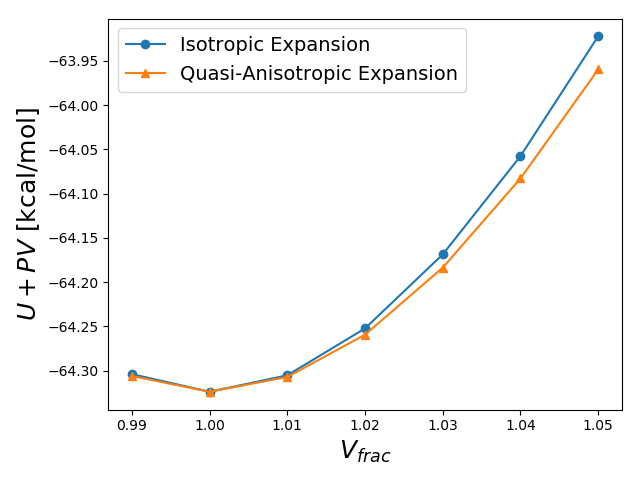}
    \includegraphics[width=8cm]{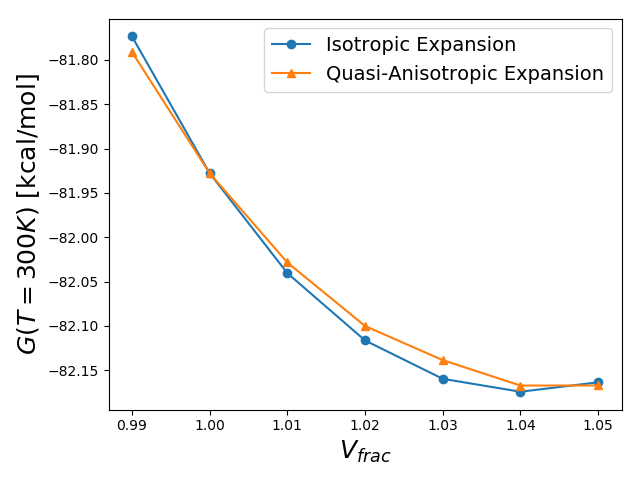}
    \end{center}
    \caption{Lattice enthalpy (right) and Gibbs free energy at 300 K for piracetam form III for varying volume fractions
             using isotropic expansion and quasi-anisotropic expansion.
    \label{figure:Quasi-Aniso_GvsV_bismev3}}
  \end{figure*}

\begin{figure*}[!htb]
    \begin{center}
    \includegraphics[width=8cm]{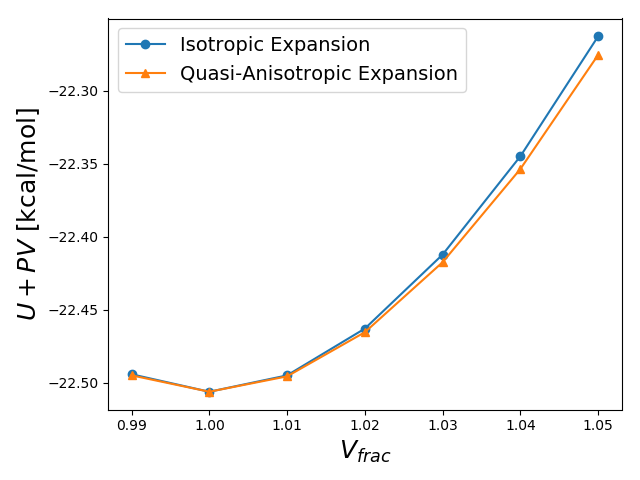}
    \includegraphics[width=8cm]{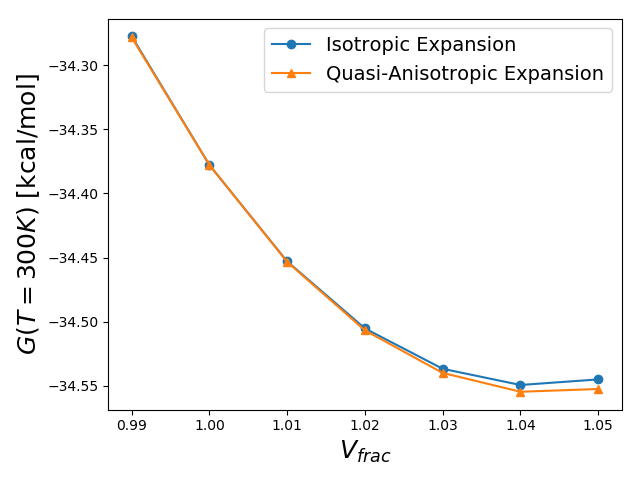}
    \end{center}
    \caption{Lattice enthalpy (right) and Gibbs free energy at 300 K for resorcinol form $\alpha$ for varying volume fractions
             using isotropic expansion and quasi-anisotropic expansion.
    \label{figure:Quasi-Aniso_GvsV_resora1}}
  \end{figure*}

\begin{figure*}[!htb]
    \begin{center}
    \includegraphics[width=8cm]{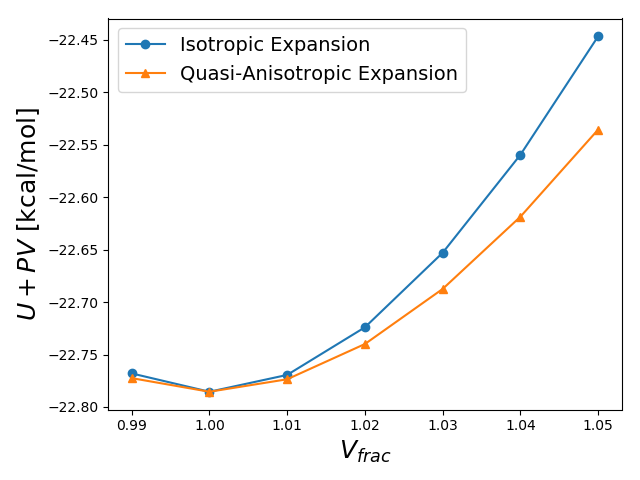}
    \includegraphics[width=8cm]{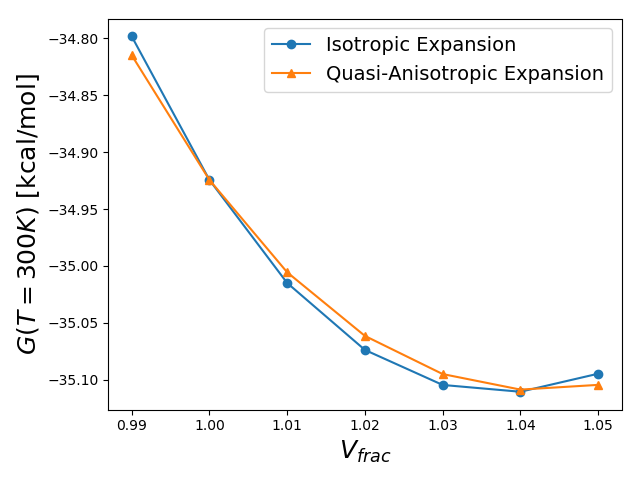}
    \end{center}
    \caption{Lattice enthalpy (right) and Gibbs free energy at 300 K for resorcinol form $\beta$ for varying volume fractions
             using isotropic expansion and quasi-anisotropic expansion.
    \label{figure:Quasi-Aniso_GvsV_resora2}}
  \end{figure*}

    Quasi-anisotropic QHA lowers the structures potential energy curve, $U(V)$, relative to isotropic expansion.
However, the Gibbs free energy, $G(T > 0 K, V)$, is not always lower. We found that the free energy
for isotropic expansion could be lower in energy than the free energy curve found by quasi-anisotropic expansion, as the decrease in lattice energy compared to the isotropic case is compensated by an increase in vibrational energy.
For all systems, we compared the free energies versus volume at 0 K and 300 K for the isotropic and quasi-anisotropic 
structures at varying volume fractions. Figures ~\ref{figure:Quasi-Aniso_GvsV_bismev1}--\ref{figure:Quasi-Aniso_GvsV_resora2} 
shows the $U(V) + PV$ curves at 0 K (left) and the $G(V)$ curves at 300 K (right). At all volumes, except the lattice 
minimum structure, the constrained optimization properly reduces the potential energy by 0.0--0.1 kcal/mol relative to the
isotropic structures. The figures on the right qualitatively show that the isotropic structures in many cases have lower free
energy than the quasi-anisotropic structures. 

\section{Modifications to Tinker} \label{section:Mod_Tinker}
    All work in this paper was completed in Tinker 8.1 molecular modeling Package with modified code. We cannot 
re-distribute these modification, but comment on the changes made. All modifications were to improve the accuracy and
physical validity of the {\tt vibrate} executable, which computes and returns the eigenvalues of the Hessian and 
mass-weighted Hessian. The eigenvalues of the mass-weighted Hessian are the crystals vibrational spectra in the form
of wavenumbers.

    Several of our systems from our previous work were outputting large negative wavenumbers, which
is non-physical. For a properly geometry optimized structure, there should $3N-3$ positive wavenumbers and $3$ zero 
wavenumbers, where $N$ is the number of atoms per cell. In the source code 'echarge2.f', which calculates the second 
derivative of the charge interactions, there is a comment noting that the reciprocal space contribution is neglected.
We have implemented a central finite difference of the second derivative, using the correctly calculated first derivatives
of the charge energy, to include the contribution to the reciprocal space. We verified the new vibrational spectra
with {\tt testhess}, a numeric solution of the Hessian, and have verified that the wavenumbers are all non-negative
and have better agreement with the numeric solution than the previous analytic solution had.

\section{Computing the Strain for the Gr\"{u}neisen Parameter} \label{section:Computing_Strain}
   In the main paper we discussed how we computed the strain, $\boldsymbol{\eta}$, between a deformed crystal 
$\boldsymbol{C}$ and a reference crystal $\boldsymbol{C_{0}}$. Here, we discuss the derivation to get the equations in
the main paper and how we remove any artificial rotation in present in the crystal.

\subsection{Strain Treatment}
    For our $3\times3$ symmetric strain tensor, $\boldsymbol{\eta}$, we can compute a deformation gradient that is 
applied to the crystal lattice vectors. Since the strains applied to the lattice minimum crystal to minimize the
free energy at $T > 0K$ are small, we have chosen to use the small strain definition to relate our desired strain
with the deformation matrix, $\boldsymbol{F}$.
  \begin{eqnarray}
    \boldsymbol{\eta} = \frac{1}{2}(\boldsymbol{F} + \boldsymbol{F}^{T}) - \boldsymbol{I}
  \end{eqnarray}
We assume that angular momentum is conserved in the crystal matrix (i. e. no rotational component of $\boldsymbol{F}$),
which implies that $\boldsymbol{F}$ is symmetric and $\boldsymbol{F} = \boldsymbol{F}^{T}$. Our relationship between
$\boldsymbol{\eta}$ and $\boldsymbol{F}$ reduces to:
  \begin{eqnarray}
    \boldsymbol{\eta} &=& \boldsymbol{F} - \boldsymbol{I} \\
    \boldsymbol{\eta} + \boldsymbol{I} &=& \boldsymbol{F}
  \end{eqnarray}

    For a given crystal we can define the crystal tensor $\boldsymbol{C}$ that is composed of the three vectors that 
describe the crystal space. If we wanted to determine the deformed crystal tensor, $\boldsymbol{C^\prime}$, of a crystal 
deformed by strain $\boldsymbol{\eta}$, we would use the following relationship:
  \begin{eqnarray}
    \boldsymbol{C^{\prime}} &=& \boldsymbol{F} \boldsymbol{C} \\
    \boldsymbol{C^{\prime}} &=& (\boldsymbol{\eta} + \boldsymbol{I}) \boldsymbol{C}
  \end{eqnarray}
We can then re-arrange this to get resulting strain:
  \begin{eqnarray}
    \boldsymbol{\eta} = \boldsymbol{C} \boldsymbol{C_{0}}^{-1} - \boldsymbol{I}
    \label{equation:strain}
  \end{eqnarray}

\subsection{Removing Artificial Rotation}
    There is an artificial rotation in the computation of the strain in eq.~\ref{equation:strain} that must be removed
before the strains can be used for the Gr\"{u}neisen parameter. In eq.~\ref{equation:strain}, both crystal tensors are
upper triangle matrices, which will lead to an asymmetric strain tensor. Since we are working with a bulk material in
periodic boundary conditions, the rotation of the vectors in space will have zero effect on the potential or vibrational
energy of the crystal. This artificial rotation between the two crystal tensors, $\boldsymbol{C}$ and 
$\boldsymbol{C_{0}}$, must be removed so we can compute the true strain between the two crystals. We use a decomposition
method by Hoger and Carlson to determine the rotation and true strain applied to the 
crystal.~\cite{hoger_determination_1984}

\section{Matching Vibrational Modes} \label{section:matching_modes}
    For all Gr\"{u}neisen and the quasi-anisotropic approaches we are fitting each vibrational mode to a functional form.
TINKER outputs the frequencies in numerical order, which leaves the possibility of mismatching between modes between two
different structures. To assure that modes match before fitting our functional form we determine a weight between all 
eigenvectors and then use the Munkres python package to match each mode.

\section{Constrained Volume Optimization \\Procedure} \label{section:constrained_optimization}
    In the semi-anisotropic QHA method the lattice minimum structure is compressed and expanded to several volume fractions
and then each expanded structure is optimized while constraining the volume. We use the {\tt scipy.optimize.minimize}
to perform the constant volume lattice minimization. We used the 'SLSQP' method, $tol=None$, $ftol=10^{-6}$, and $eps=10^{-4}$
all of which were optimized for computational speed and convergence within $10^{-8}$ kcal/mol.

    To assure that the structure is at a minimum we used the method of Lagrange multipliers, where the Lagrange function 
to be minimized is shown in eq.~\ref{equation:Lagrange_func}.
  \begin{eqnarray}
    \mathcal{L} = U(\boldsymbol{C}) - \lambda (V_0 - V(\boldsymbol{C}))
    \label{equation:Lagrange_func}
  \end{eqnarray}
The derivative of $\mathcal{L}$ with respect to each $C_{i}$ ($i=1,\ldots,6$) and $\lambda$ should be zero when the crystal
structure is optimized to the energy minimum at a given volume ($\mathcal{L}^\prime = 0$). Due to numerical error, there 
values of $\mathcal{L}^{\prime}$ may be slightly non-zero. We checked the percentage of the derivatives of the
optimized structure relative to the isotropic structure, shown in eq.~\ref{equation:Lagrange_frac}.
  \begin{eqnarray}
    \frac{\mathcal{L}^{\prime}_{aniso.}}{\mathcal{L}^{\prime}_{iso.}} \times 100\%
    \label{equation:Lagrange_frac}
  \end{eqnarray}
At 100\% the structure is completely isotropic and 0\% would be the lattice minimum structure for that volume. We found
that all lattice parameters were $< 3\%$ for all optimized structures and the average value was 0.25\%, which is within 
the range of accuracy for the results presented in this study.

\newpage
\newpage


\newpage
\newpage

\bibliography{citations}

\end{document}